\documentclass{article}[a4paper]

\title{ Humanoid Robots at work: where are we ?}
\author{Fabrice R. Noreils \thanks{The author got his PhD on robotics and Artificial Intelligence at LAAS CNRS, Toulouse France. He was a researcher on task planning and Execution on wheeled exploratory robots and led R\&D teams on resources optimization for a LEO constellation of satellites, character animation of small humanoid robots, autonomous navigation of an indoor flying drone and task planning and navigation of a fleet of AMRs in logistics environment. He had operational roles such as the creation and management of a team of field engineers to deploy AMRs and a team of operators to remotely supervise them. The author is now a technical leader for L3/L4 highway automated vehicles. The opinions expressed in this publication are those of the author. They do not purport to reflect the opinions or views of his current/former employers.}}

\usepackage{graphicx}
\graphicspath{{./images/}}
\DeclareGraphicsExtensions{.jpg,.png}
\usepackage{amsmath}
\usepackage{leftidx}
\usepackage{url}
\usepackage{tocloft}
\usepackage{enumitem}
\usepackage{booktabs}
\usepackage{geometry}
\usepackage{wrapfig}
\usepackage{multicol}
\usepackage[
bookmarksopen,
bookmarksdepth=2,
breaklinks=true
]{hyperref}
\usepackage[anythingbreaks]{breakurl}
\hypersetup{
   colorlinks   = true, 
   urlcolor     = blue, 
   linkcolor    = blue, 
   citecolor    = red   
}
\newgeometry{vmargin={25mm}, hmargin={20mm,17mm}} 

\begin {document}
\pagenumbering{gobble}
\maketitle
\pagenumbering{arabic}

\section{Introduction}

In 2021 Elon Musk showed off humanoid robot prototype at Tesla AI day. Musk said
robots made by Tesla could one day be worth more than its cars, and that
thousands of them would be put to work in Tesla factories, where humans build
cars and batteries.

Without knowing it, Elon Musk launched a race in which many companies have already engaged, mainly on the North American continent and in Asia. Investment funds are scrambling to invest in what they think is a new El Dorado~\cite{fortune_2024_1}.

A humanoid robot is an extremely complex machine at the intersection of many disciplines such as mechatronics, control algorithms for underactuated dynamic systems, actuators design, energy consumption optimization, autonomous system in terms of perception localization planning and locomotion...

In addition, all algorithms must be carried out by on-board computers in a very constraint footprint.

Finally, the objective is to put these very complex, therefore fragile, machines in a demanding industrial environment to perform at least as well as a human worker within 2 or 3 years?

This is reminiscent of the enthusiasm and certainties of promoters of autonomous vehicles in the 2010s.

The question we can ask ourselves is the following: is this objective realistic or is it achievable not in 2 or 3 years but within 10 years or more? The aim of this document and its main contributions is to provide some hints by covering the following topics:

\begin{enumerate}
	\item An analysis of 12 companies based on eight criteria described in section~\ref{competitors_analysis}. Although these criteria are subjective, they have the benefit of being able to distinguish companies based on their maturity and approach to the market;
	\item These humanoids are very complex systems facing technological challenges which are discussed in  section~\ref{technological_challenges};
	\item Operation and maintenance are critical activities specially when humanoids are deployed at scale. This topic is covered in section~\ref{support_maintenance};
	\item Pilots are the last step to test the feasibility of a new system before mass deployment. This is an important step to test the maturity of a product and the strategy of the humanoid supplier to address a market. This topic is addressed in section~\ref{deployment}.
\end{enumerate}

\section{General Considerations}

In this section, contextual considerations are provided.

\subsection{Humanoid Robots because Shortage of labor, really?}

Most of the companies working on commercial humanoids, sees warehouses as an
obvious entry point because the recurrent argument from the logistics industry is: shortage of labor. 

Let us have a look at the HSE (Health and Safety Executive) report on Transportation and Storage statistics in Great Britain, 2022~\cite{HSETechreport}:
\begin{itemize}
\item 49 000 workers suffering from work-related ill health: musculoskeletal
disorders (36\%), stress, depression or anxiety (41\%) and other illness (23\%),
\item 16 fatal injuries to workers – broadly similar to the five-year average of
14 fatalities per year – of which 34\% were due to being struck by a moving
vehicle, 21\% resulted from a fall from height, and 11\% involved being struck
by a moving/falling object,
\item 31 000 non-fatal injuries: slip, trip or fall (32\%), lifting/carrying
(23\%) and being struck by a moving/falling object (13\%).
\end{itemize}

Two more data to complete the overall picture:
\begin{itemize}
	\item The median annual earnings in the United Kingdom was 34,963 British pounds per year in 2023. 
	\item A recent survey found that the average turnover rate in the logistics industry is around 31\%.
\end{itemize}

What do these numbers tell us? For a relatively low salary, workers are dealing with a harsh environment and have a high chance of getting hurt over a period of time of only a few years, and this is why turnover is very high. This forces companies to spend money on recruitment, training and medical aid. 

Saying that there is a shortage of labors is a fallacious argument because there are people willing to work but not at any condition. Besides, the logistics industry can rely on immigration to hire new workers as well. 

An obvious solution consists in increasing salaries and improving working conditions, but it will hurt margins. This is probably why the logistics industry turns towards automation and robotics. Robotics is covering a wide area of solutions from product storage, AMRs, manipulator arms and ... the newcomer which is humanoid robots.

Regarding humanoid robots, two other pieces of information deserve to be cited to complete the picture:

\begin{itemize}
\item The maximum recommended weight limit for human workers under the National Institute for Occupational Safety and 
Health’s Lifting Equation is 51 lb (23 kg).  UPS requires labeling and special procedures when shipping packages over 70 lb (31.5 kg).  Implementation of the Manual Handling Directive in the European Union varies by member state,  but several countries specify a 25 kg limit for men and 15 kg limit for women. These weight limits provide a general guideline for the dynamic loads that most humanoid robots should be expected to carry.

\item As mentioned above, the median annual earnings in the United Kingdom for a worker in a warehouse was 34,963 British pounds per year in 2023 (around US\$ 44k). It is not a surprised if many companies working on humanoids are targeting a price below US\$ 50k or even lower. 
\end{itemize}

\subsection{After AI, China is betting on humanoid robots}

China created a list of key technologies in which they want to be world-leader. In 2017, China has outlined plans to become a world-leader in artificial intelligence by 2030 ~\cite{ChinaLeaderAI}. On November 2nd, 2023, the Ministry of Industry and Information Technology (MIIT), published a nine-page guideline on its website, saying that China should realize mass production of humanoid robots by 2025 and  humanoid robots should become an important new engine of economic growth by 2027\cite{ChinaLeaderHumanoidRobots}.

In August 2023, the World Robot Conference (WRC) held in Beijing was the China’s largest robot exhibition~\cite{WRC2023}. More than 160 companies were represented and more than 600 robots were exhibited to the public. The public was astonished by the humanoid robots  with, for some of them, a more and more realistic human appearance. Clearly, it was an opportunity for the Chinese government to show the level of technology reached by some of its local companies.

A strong signal has been sent to the Chinese R\&D robotics community: the
central government will pour a lot of money in humanoids. Therefore, it is likely we will witness a wave of 
humanoid supplier companies in the years to come and a fierce competition will take place so that 2 or 3 majors emerge to address the market worldwide and few other dedicated  to the domestic market. We will have to wait until 2025/2026 to
know the winners.

\subsection{What kind of business model for humanoid robots?}

These humanoid robots will be expensive, even though they will be mass product. The question then is how to sell them? One answer may be to look at the Autonomous Mobile Robot (AMR) market. Indeed, there’s lot of companies that are selling mobile robots to these exact same markets like logistics which are targeted by humanoid supplier companies. The most commonly business model chosen by customers is robots as a service or RaaS. This business model did not exist before the AMR market took off.

Customers at least in the early stages of the humanoid market  want to see how this works (see also section~\ref{deployment} for more details) and this is why it is likely they will go for robots as a service model.
Besides there is an eco-system with third party financing groups already set up with the AMR market on which the emerging humanoid market can rely on.

Worth to mention that the companies which are making humanoid robots need a lot of cash a different stage of their growth. Humanoid robots are very expensive to build (hardware, structural materials like aluminum or plastic, 3D printing or machining tools, Software development tools... and many highskill professionals). Therefore these companies are well funded (from US\$ 20 to 50 Millions) to develop a prototype reliable enough to initiate and finance pilots. Successfull pilots will trigger production. It means that a company will raise another US\$ 200-300 Millions to build a factory, hire staff and start the production but also develop its operation and support teams. Later on, more financing will be necessary either to cover the losses or expand the market share.

\section{Which Are The Competitors in this new race?\label{competitors_analysis}}

\subsection{Introduction}

In this section we will present the companies\footnote{For the sake of generality I will use the term company(ies) when I will refer to one or more of these humanoid manufacturer(s) in the remaining of this article, unless I mention one specifically.} which aim to commercialize a humanoid robot in the years to come. Each company develops its own strategy and approach to the market. Therefore each company will emphasize certain technologies than others. In order to provide a global picture, 8 criteria are proposed which are described below as well as a radar chart for each of them. The description of each company according to these criteria is given in Appendix~\ref{menagerie}.

We are considering the following criteria:

\begin{itemize}
\item \textbf{Robotics background}: it characterizes the expertise and experience of the teams in charge of developments (electro-mechanics, control, localization, path planning, perception and AI). We propose  0.2: the company is young and does not communicate on the skills of its engineers, 0.5 the company hired engineers/researchers recognized by the community for their contribution to humanoid robotics 1: the company has a long experience in the development of humanoid and/or quadruped robots;

\item \textbf{Teleoperation}: teleoperation can play an important role to train a deep net how to grap objects when the robot is equiped with hands or to takeover by an operator when the robot is stucked for example. However teleoperation is not mandatory for a humanoid robot which will focus on a single task like moving totes for instance. We propose 0: teleoperation is not mentioned because the company is relying on another method, 0.5: teleoperation as a tool for training and 1: multi-purpose teleoperation (training, teleoperated to achieve a very complex task or takeover in case of failure);

\item \textbf{Modularity}: for certain tasks in a warehouse which is built on a flat floor, legs it is not a necessary asset, wheels are enough or even the customer is looking for a fixed robot capable of sorting objects on a conveyor. This means more opportinities that the company to sell its humanoid robot in parts (upper body on wheels, legs or a fixed platform). We propose 0.5: full humanoid only and 1: upper body on wheels, legs or fix platform;

\item \textbf{Dexterous hands}: these humanoid robots for the most part are intended to be general purpose robots, which means that they can work in various environments, pick up or sort different type of objects. It is therefore necessary to develop an agile hand with tactile sensors. We propose  0.2: gripper, 0.6: hand with less than 10 DoF and 1: hand with more than 10 DoF;

\item \textbf{Task planning}: it means the possibility for the robot to plan by itself a sequence of tasks from an order input (textual or vocal). Although this is normally part of AI, I am deliberately separating it because it is still a very active area of research and which will not be integrated into these robots in the short term. We propose  0: activity not disclosed or not mentioned, 0.2: mentioned by the company but not demonstrated, 0.6: mentioned and demonstrated in videos or papers and  1: included in the commercialized robot within the year to come;

\item \textbf{AI}: it's a very generic term and I use it here to differentiate different stages of AI integration. We propose 0.2: basic capabilities such as mapping, localization, perception for path planning, control, 0.4 plus objects detection and classification, 0.6 plus End-to-end Perception-grasping and 1: end-to-end from locomotion /control - perception - grasping;

\item \textbf{Walking gait}: it is obvious that a humanoid robot must walk. This is the field where one can see spectacular results but there is a difference between a humanoid which is steadily walking at 2m/s and another one which can walk at 3 or 4 m/s on an uneven terrain while an operator is pushing it. We propose 0: not seen any video, 0.6: video of a dynamic walking gait and 1: videos showing robustness with respect to kicks/pushes;

\item \textbf{Market/pilot}: all these companies want to address one or more markets. What I note here is whether there are pilots in progress, the company's willingness to describe the pilots and also the willingness to open the platform to labs or other companies to develop new applications. We proposed 0: no pilot disclosed, 0.2:open platform available, 0.5: pilots but no feedback disclosed and 1: pilot and information disclosed.

\end{itemize}

These criteria are subjective, but they have the advantage of being able to distinguish companies based on their maturity and approach to the market.

\subsection{Panorama and analysis}

\begin{figure}[!h]
	\begin{center}
	\includegraphics[scale=0.7]{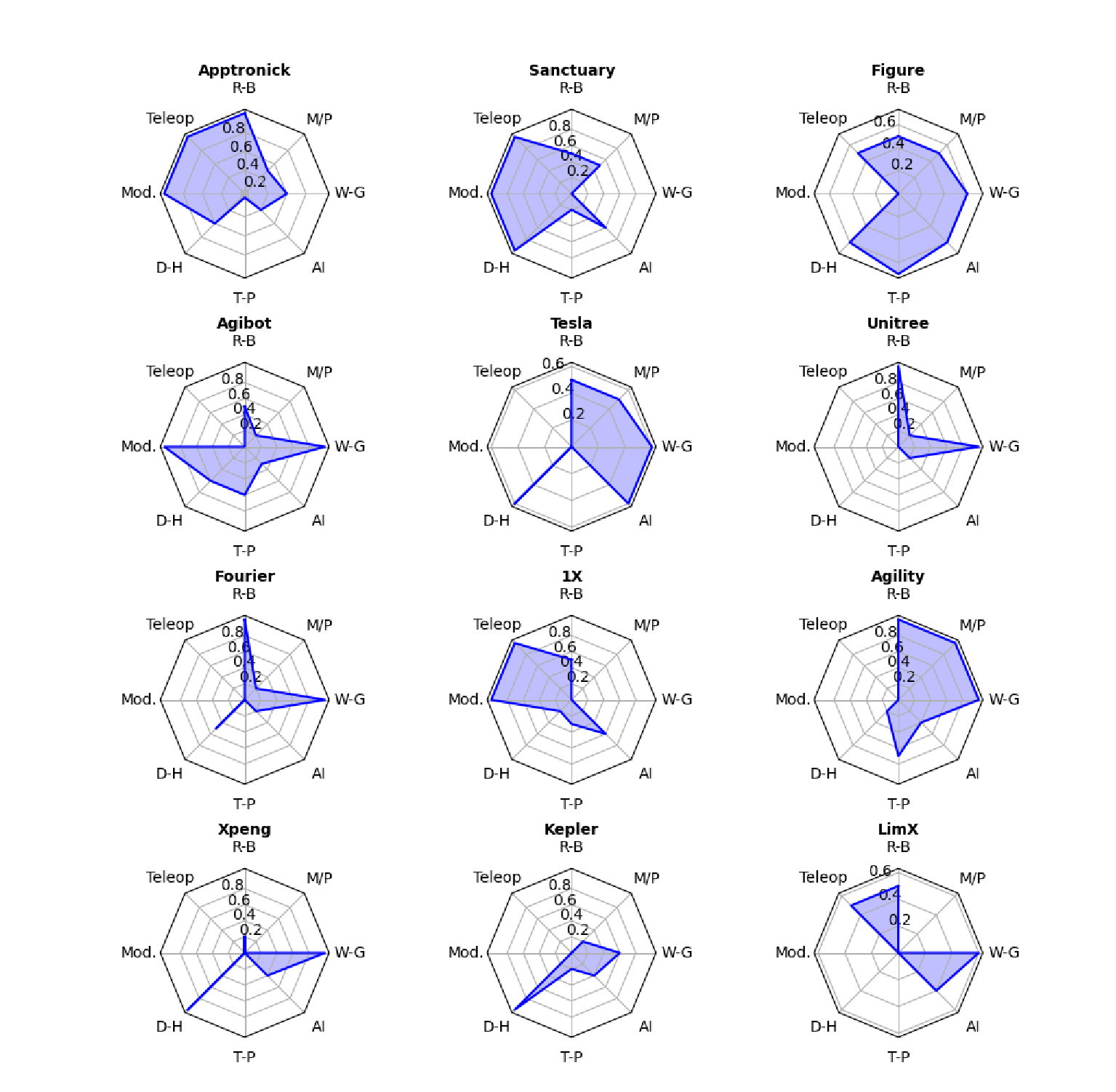}
	\caption{Radar plots of the eleven companies. R-B (Robotics Background), Teleop (Teleoperation), Mod (Modularity), D-H (Dextereous Hands), T-P (Task Planning), AI (AI), W-G (Walking Gait), M/P (Market/Pilot)}
	\label{radars}
	\end{center}
\end{figure}

The picture~\ref{radars} highlights the radar plots of the different companies we considered. This analysis is based on a photo at the time this document was written. It is very likely that within six months or one year new companies will appear or that the companies cited in this document will have improved their offer in terms of humanoids (software, mechanics, etc.).

The competition is between North America (4 companies) and China (6 companies, the central government's announcement to finance this technology is certainly not unrelated). Europe is represented by a single company. From these plots, we can outline different strategies:

\begin{itemize}

\item \textbf{platform oriented} We note that Fourier, Unitree and LimX have a fairly similar footprint, the robot is an elaborate humanoid platform  and research centers are encouraged to develop applications. Although the Unitree humanoid is equipped with electrical motors, it has exceptional performance in terms of locomotion;

\item \textbf{Modularity, teleoperation, AI, hands or legs} Santuary, Apptronik and 1X have the same strategy, teleoperation plays an essential role for learning, the modularity of the commercial offer as legs are not a must have asset - at least in a short term frame - and a significant investment in AI algorithms and task planning (there is nevertheless a notable technological difference concerning the development of the hand which is a technological marker for Santuary and under development for 1X which);

\item \textbf{Pilot ready} Agility, Figure and Tesla have a very similar approach. They have strong teams, a humanoid robot developed enough to start pilots. Note that the radar gradation goes from 0 to 1 for Agility while it goes from 0 to 0.8 for Figure.ai and Tesla. Agility Robotics, very pragmatic in its approach, is very advanced in its deployments in logistics - see Section~\ref{deployment};

\item \textbf{Very Promising and to follow} Some companies like Agibot are very young (less than a year) but with a very elaborate long-term vision and a high-performance biped robot very similar to the one from Agility Robotics (see Table~\ref{agibot_table}). This is definitely a company to follow. The same goes for Kepler which showed his humanoid at CES 2024.

\end{itemize}	


I focus on companies that are developing a humanoid robot with the aim of commercializing it in the next 2 or 3 years.This is why I did not include Boston Dynamics' Atlas, the world's most advanced humanoid robot, at least in terms of locomotion, because on one hand it is a very expensive piece of engineering and on the other hand, it is an R\&D platform use to explore new algorithms on dynamic whole-body control. However Boston Dynamics may consider Atlas as a product in the future as it  released a \href{https://www.youtube.com/watch?v=SFKM-Rxiqzg}{video} showing its Atlas humanoid picking and placing automotive struts. The significance of the demonstration is that Atlas performs all of the object recognition using the robot’s onboard sensors. Atlas acquires the automotive struts, using its grippers from a vertical storage unit, and places them horizontally onto a flow cart~\cite{robotreport_atlas_2024}.

There are many very successful companies in Europe in the
humanoid robotics field like \href{https://pal-robotics.com/}{Pal Robotics} which develops humanoid robots intended for research laboratories such as \href{https://pal-robotics.com/robots/talos/}{TALOS} or \href{https://pal-robotics.com/robots/kangaroo/}{KANGAROO}.

We must also mention \href{https://enchanted.tools/}{Enchanted Tools} with his Mirokai robot which consists in humanoid upper body on a ballbot. I did not consider it in the table because all the other robots have legs. However, We will talk about it again in the section related to deployment because the company's approach is very pragmatic.

\section{Technical Challenges \label{technological_challenges}}

This section aims at providing a panorama of the advances in different fields that foster the development of humanoid robots.

\subsection{Mechanics/actuators/control}

\begin{figure}[!h]
	\centering
	\includegraphics[scale=0.3]{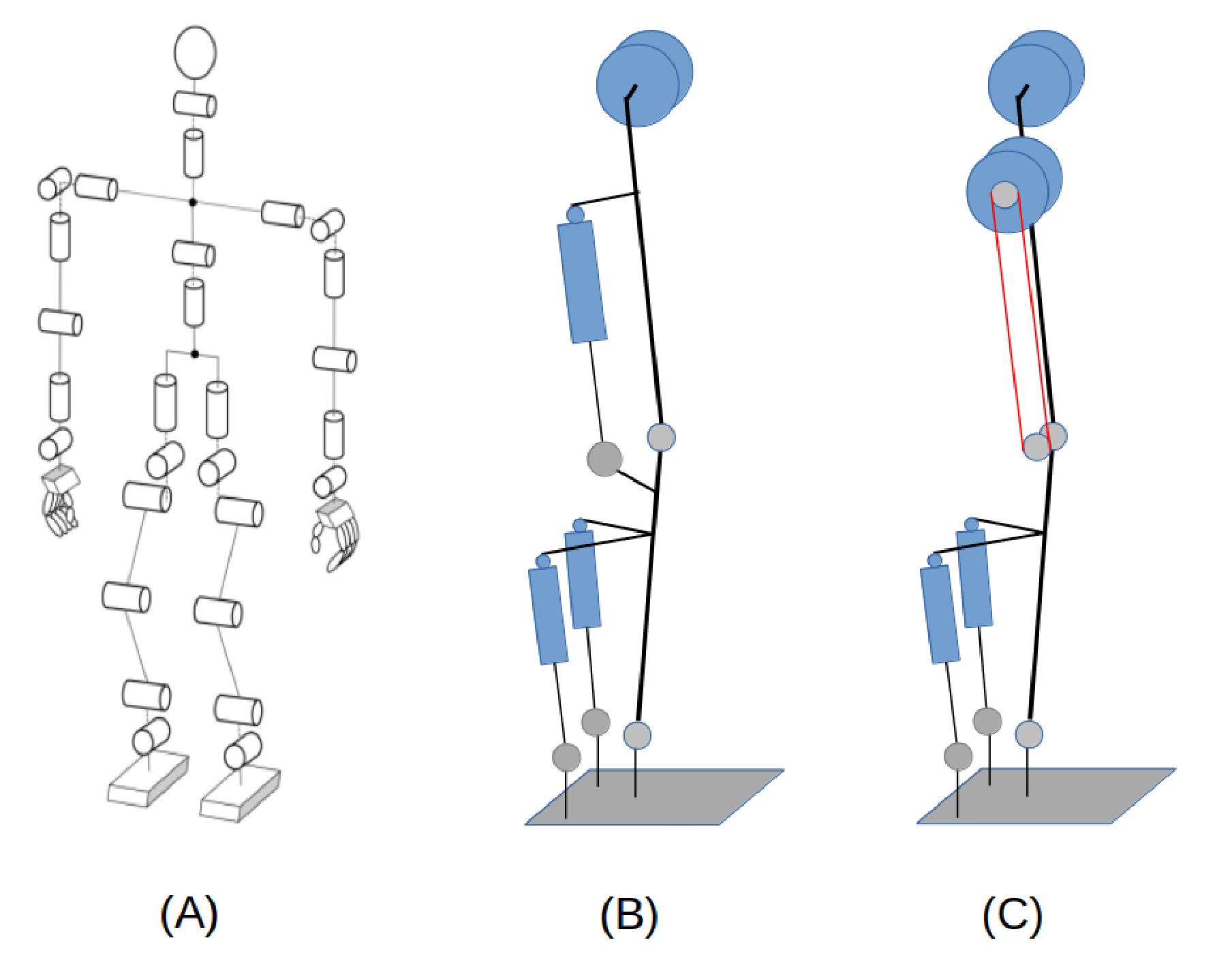}
	\caption{From a classical wellknow serial configuration (A) to Linear parallel mechanism (B) or combination of serial and parallel mechanisms (C)}
	\label{fig:configurations}
\end{figure}

For a long time humanoids were similar in terms of kinematics. The rotary joints are powered with servo-motors in a serial
configuration (Figure~\ref{fig:configurations} (A) from ~\cite{Ali2010ClosedformIK}). The structure has been thoroughly investigated and  a closed-form solution for the inverse kinematics can be derived which is one of its main benefit. Unfortunately there are several drawbacks as the rotary actuators are usually located directly at the joint, the ones closer to the origin of the chain need to carry the ones lower in the chain. Position errors are summed across the joints which decreases accuracy. A serial architecture also has higher inertias and therefore lower accelerations. It is worth noting that H1 from Unitree is the only one humanoid which keep this serial configuration.

The experience and the theory shows that a low-mass, low-inertia leg with high bandwidth actuators allow for many control simplifications. And this is what we observe on this new generation of humanoids. Indeed a solution consists in attaching the actuators off-axis, closest to the root of the link and using a lightweight coupling or transmission like a belt - see Figure~\ref{fig:configurations} (B) in which the rotary actuator responsible of the knee is close to the hip in order to reduce the leg inertia~\cite{Ficht2021BipedalHH}. As the tigh and the shank usually provide enough space for housing  actuators and couplings, several efficient solutions have been developed such as crank-lever mechanims and linear actuators see Figure~\ref{fig:configurations} (C). Almost all the new humanoids using two linear actuators for pitching and
rolling the ankle around a Cardan joint.

In terms of actuators, one can notice that all companies designed and developped their own electrical rotatary and linear actuators. These Humanoid robots will interact with their environment and they need to absorb impacts while walking, force-feedback control is mandatory. 

The development of quadruped robots was an opportunity to design a generation of actuators referred to as quasi direct
drives ~\cite{Kalouche2017GOATAL} ~\cite{Katz2019MiniCA} which have been developed to generate sufficiently large torques, without sacrificing back-drivability and/or mechanical resilience. These drives are often composed of torque motors (in contrast to power/speed motors) and single-stage planetary gearing transmission which is lower than 1:10, so that
the output torque is amplified to some extent and intrinsic back-drivability is achieved, without
compromising the system bandwidth. Moreover, the low reduction of the gearing and the resulting high efficiency allows the motor torque, that can be realized by the current measurement, to be directly a valid indication of the output torque. Larger version of these quasi direct drive actuators are used on the humanoid robots.

The new humanoids are also relying heavily on linear actuators. I cannot tell whether they are mechanical linear motor (designed around ball screw drives and backdrivability is disputable) or direct drive linear actuators. Direct drive linear motor consists of two parts: the slider and the stator. The slider is made of magnet while the stator contains the motor windings. It has been shown that this kind of motor is highly backdrivable and suitable for humanoid robots~\cite{Lucidarme2019PreliminarySO}.

In terms of control, if reinforcement learning lead to amazing results for quadrupeds, "newtonian" optimization is mainly applied to humanoid robots~\cite{Wensing2022OptimizationBasedCF}. Moreover over the years, open source  control toolboxes have been developed such as DRAKE~\cite{drake}, pinocchio~\cite{carpentier2019pinocchio} or IHMC~\cite{IHMC} to name a few. These toolboxes synthesize the "know-how" of a decade of research  on dynamic walking. New researchers can leverage on them to quickly prototype a new algorithm and test it on a humanoid.

\subsection{End-to-end grasping with dextereous hands\label{end-to-end-grasping}}

This the robotic manipulation golden age! Indeed the number of publications investigating AI approaches for 6-DoF grasping have grown significantly in the last few years and the achievements were almost not thinkable 5 years ago. It was possible because we observed amazing progress in different connected fields as well. Let us have a look.

General purpose humanoid means that this robot will work and manipulate the same objects as human workers do. Therefore a dextereous multi-fingered hand will be meaningfull. The design and the development of this kind of hand is a challenge by itself: the mechanical part to ensure high dexterity, intrinsic and extrinsic sensors and the grasping force control. However, with the rapid development of manufacturing and processing technology, new materials and electronic, mechanical are emerging, which could provide new structures, actuators, and sensors for multifingered hand design~\cite{li_frontier_2022}~\cite{Xia2022ARO}.

Simulation softwares like deep net algorithms benefit from GPU advances.  Game-engine renderers can model cameras well enough not only to test a perception system in simulation, but even to train perception systems in simulation and expect them to work in the real world! There is also an amazing improvements in the quality and performance of contact simulation. Making robust and performant simulations of multi-body contact involves dealing with complex geometry queries and stiff (measure-) differential equations. These simulation softwares allow for the fast development of new algorithms and/or training new policies ~\cite{manipulation_drake}.


Once you have designed a hand, you are halfway there, then you need to know how to recognize the object you want to grab and how to pick it up. This is a very hard problem, and again the recent  advances in AI will give us the tools to solve it. In perception to grasping arena, imitation learning is heavily used to train a neural net how to grasp a given object according to its shape and softness.  Different techniques have been explored such as vision based teleoperation ~\cite{Li2018VisionbasedTO} ~\cite{Handa2019DexPilotVT} or learning dexterity from internet videos~\cite{Shaw2022VideoDexLD} for instance. The benefit of these techniques is their low cost but they lack of haptic and forces/torques feedback. Some teleoperation equipments are able to provide these feedbacks but they may be expensive to acquire at large scales. 

Another approach which seems very promising is Diffusion Policy~\cite{Chi2023DiffusionPV} which enables easy and rapid behavior teaching from demonstration. Toyota Research Institute (TRI) leverages on diffusion policy and developed a pipeline such that  TRI’s robot behavior model learns from haptic demonstrations from a teacher, combined with a language description of the goal.  A proprietary AI-based Diffusion Policy is used to learn the demonstrated skill. This process allows a new behavior to be deployed autonomously from dozens of demonstrations~\cite{TRI_2023}. Diffusion model is also the foundation of new research aiming at developing a policy composition framework domains (simulation, real robot of video) and modalities (vision, depth images, tactile...)~\cite{Wang2024PoCoPC}. 

Researchers are also developing tools to simplify as much as possible the collection of data. For instance this paper~\cite{chi2024universal} introduces what they call a Universal Manipulation Interface (UMI) -- a data collection and policy learning framework that allows direct skill transfer from in-the-wild human demonstrations to deployable robot policies. In-the-wild because it employs hand-held grippers coupled with careful interface design to enable portable, low-cost, and information-rich data collection. It is quite a breakthrough in the senses that it eases significantly the training of a robot policy for a given task.

To conclude this brief overview, so far all these humanoid robots are manipulating small and light objects and thus the lower part of their body is rigid. However if they have to manipulate bulky objects they will have to engage their entire body like a human worker to avoid injuries and for the robot to avoid actuators to be destroyed too quickly. Some researchers are staring to work on this field as well ~\cite{Seo2023DeepIL}.

\subsection{Spatial reasoning and tasks planning (for manipulation)\label{spatial_reasoning}}

We can start by mentioning two videos that have been posted recently:

\begin{itemize}
	\item Agility Robotics, in a \href{https://www.youtube.com/watch?v=CnkM0AecxYA}{video}, demonstrated that it is possible to give verbal orders to digit which was able to plan a sequence of tasks to execute the order by using a dedicated LLM.  Although the environment is very simple, a set of towers of different heights as well as three boxes, each one with a a different color and pictograms, and the orders like "move the red box one the lowest tower" for instance, digit was able to execute the order; 

	\item Figure, which partnered with OpenAI, released a \href{https://www.youtube.com/watch?v=Sq1QZB5baNw}{demo video} where the robot can talk like a human. The robot was able to identify things put in front of it, answer the queries, do the task asked of it (giving an apple to the person), and explain how it did that at the same time it was doing something else.
\end{itemize}

These demos were possible with the recent advances in what we can call \emph{Robotic Transformers} and different approaches are explored by researchers.

One approach consists in combining a Large Language Model (LLM) and Vision Language Model (VLM).  LLMs pretrained on broad web-scale datasets have demonstrated reasoning skills over text, leveraging their natural language cpabilities and common-sense reasoning for generating robot tasks. However LLMs can generate plans but they cannot perceive the environment. At the same time, VLMs enable open-vocabulary visual recognition and are able to make complex inferences about object-agent interactions in images. A VLM, given an image, is able to perform spatial reasoning and object classification. Based on these information a LLM is able to plan a sequence of actions corresponding to a specific query. Palm-E is an example of this approach ~\cite{Driess2023PaLMEAE}. Note that the sequence of actions is then executed by low level controllers. For old researchers, it may recall them STRIPS~\cite{Fikes1971STRIPSAN} the planner created for Shakey by the SRI.

The aforementioned approach does not have knowledge about the robot and controller capabilities. Some researchers argue that grounding the language belief with visual and motor feedback will lead to better performances.Therefore another approach is trying to map natural task description to robot actions. One example is RT-2, referred to as Vision-Language-Action (VLA) model, which consists in co-training the model on internet-scale data with images and actions related tasks. The position, orientation and elevation of the end effector of the robot are tokenized like text~\cite{Brohan2023RT2VM}. Note that a VLA leverages instantiations of the previous LLM+VLM models plus fine tuning these models with robot actions. 

In the section~\ref{end-to-end-grasping} I mentioned the amazing results obtained with diffusion policies. Indeed these policies are remarquably robust to external disturbances and fast to learn - like 50 training by an expert may be enough. Another approach is looking for generalist robot policies (GRPs) with the belief that these generalist robot policies have the potential to transform how robot learning research is done: in the same way that current models in NLP are almost universally derived from pretrained large language models, future robot policies might be initialized from GRPs and finetuned with modest amounts of data. This is the path followed by TRI ~\cite{TRI_2023} and an open source initiative referred to as Octo ~\cite{octo_2023}. It will then be coupled with a LLM to plan a sequence of tasks.


Research on using LLM to enable robots to plan a sequence of (very) simple tasks to execute verbal orders is just beginning. this section emphasizes some promising approaches related to manipulation and for a quite complete overview of robotics transformers, the reader may refer to ~\cite{Zeng2023LargeLM}.

However, there are different challenges that need to be addressed by these robotics tranformers to be widely adopted as reliable tasks planner for manipulation and a general purpose tasks planner:
\begin{itemize}
	\item  Long-horizon, multi-stage task planning requires reasoning over extended periods, which is a challenge for LLMs. To mitigate this issue, prompting schemes Chain-of-Thoughts distil complex problems into intermediate reasoning steps with
	the aid of exemplars, facilitating effective reasoning~\cite{Wei2022ChainOT};
	
	\item Tackling large-scale problems often leads to issues like hallucination or failing to consider important details, rendering their plans ineffective or error-ridden; 

	\item Working with multi-step tasks in robotics also involves dealing with uncertainties and changes in the environment which may lead at a given time to the failure of a task and trigger a replanning process. This a wellknow problem for those who address tasks planning in the pre-transformer era;

	\item The size of modern VLMs can reach tens or hundreds of billions of parameters and it is impossible to directly run such models on embedded GPUs available on the current humanoid robots - and this statement will last for years. The current solution in order to  enable efficient real-time inference consists in deploying these VLMs in a multi-TPU cloud service and querying this service over the network. Consequences will be addressed in Section ~\ref{support_maintenance}.  
\end{itemize}


\subsection{Safety \label{safety}}

Safety is mentioned in this section because it is a challenge for humanoid robots which is not really addressed yet.

In order for both AGVs and AMRs to not harm people or damage the surrounding infrastructure safety standards have been defined. AGVs and AMRs are equipped with safety sensors that will prevent these devices from contacting humans and objects. The AGV Safety Standard for the United States is referenced as
\href{https://webstore.ansi.org/standards/ansi/ansiitsdfb562019-2388609}{ANSI/ITSDF
B56.5-2019}, and for the European Union it is referenced as 
\href{https://webstore.ansi.org/standards/iso/iso36912023-2502304?source=cj&program=cj&cjevent=38d32672a8ac11ee82ee766e0a18b8fc}{EU EN ISO 3691-4:2020}

To summarize, the key risk mitigation measures necessary for an automatic guided vehicle are
divided into two principal categories:
\begin{itemize}
    \item Active Risk mitigation measures;
    \item Passive Risk mitigation measures.
\end{itemize}

The key AGV active risk mitigation measures are:
\begin{itemize}
    \item Safety laser scanner with collision avoidance system;
    \item Pressure-sensitive bumper guards;
    \item Safety PLC.
\end{itemize}
The key important passive risk mitigation measures are:
\begin{itemize}
    \item Emergency stop mechanisms;
    \item Warning lights – flashing and rotating lights;
    \item Audible warnings/ klaxon alarm signals;
    \item Awareness and safety signage on AGVs.
\end{itemize}

It is too early to know if part or totality of these measures will be applied to humanoid robots. However, there is a unique feature that humanoid robots have compared to AGV and AMR: \textbf{they can fall}.

As soon as humanoid robots will be able to interact with humans in the workplace, it will be mandatory for the humanoids to fall safely. It means:
\begin{itemize}
    \item Do no harm humans and damage the surrounding infrastructure;
    \item Do not damage the humanoid itself.
\end{itemize}

To quote Jerry Pratt~\cite{IEEEspectrum_Figure}:
\begin{quote}
    it’s critical to fall safely, to survive a fall, and be able to
get back up. People fall—not very often, but they do—and they get back up. And
there will be times in almost any application where the robot falls for one
reason or another and we’re going to have to just accept that. I often tell
people working on humanoids to build in a fall behavior. If the robot’s not
falling, make it fall! Because if you’re trying to make the robot so that it can
never fall, it’s just too hard of a problem, and it’s going to fall anyway, and
then it’ll be dangerous.

I think falling can be done safely. As long as computers are still in control of
the hardware, you can do very graceful, judo-style falls. You should be able to
detect where people are if you are falling, and fall away from them. So, I think
we can make these robots relatively safe. The hardest part of falling, I think,
is protecting your hands so they don’t break as you’re falling. But it’s
definitely not an insurmountable problem.
\end{quote}

Safety is not really addressed by humanoid suppliers yet but will be required if humanoids are allowed to work with human workers.

\section{Operation and Maintenance\label{support_maintenance}}

Why talking about operation and maintenance? Humanoid robots are very complex, fragile machines, and there are no statistics available on the breakdowns and their frequencies that may occur. Moreover as soon as they will be deployed in warehouses, companies will have to roll out all the necessary workforce and infrastructure to guarantee that the humanoids are fully operational and deliver the work in accordance with contracts signed with customers.

\subsection{Field engineers}

As soon as the company is selling and deploying humanoid robot on a customer site, operation and maintenance workforces specially trained will be involved. If the company is going for a Robotics as a Service (RaaS) business model and  it scales their humanoid fleets, support staff will need to do the same. Each customer, no matter the number of robots, will run into errors with their robots that will require human support from the RaaS provider. Let us name a few:
\begin{itemize}	
	
	\item Considering dextereous hand, if a hand is designed with cable-driven actuators, it is known that cable drives typically are not as durable and can be more finicky to keep calibrated;
	
	\item Motors are also quite fragile, as overloading them or operating in sensitive environments (dirt or dust which is common in logistics worplaces) can quickly lead to them being destroyed;
	
	\item  Force/torque (F/T) sensors issues: in locomotion, the stability assessment is done using an estimation of the Center of Pressure (CoP) obtained by F/T sensors. It means the sensors must be robust enough to not only carry the weight of the robot but also endure ground impact forces during walking. The drawback of currently available F/T sensors is that the measurements are done using transducers (e.g. strain gauges), with a direct coupling of tension induced by forces going through the sensor. It leads to long-term deformations which requires recalibration in order to prolong the usability of a sensor~\cite{Ficht2021BipedalHH}.
\end{itemize}

Indeed, most of the time a subscription-based model for robots means customers don’t own the robot and thus all costs of the maintenance of the hardware falls to the RaaS provider. Therefore the company will have to hire and train these staff members that will grow as the company does. 

\subsection{Integration with existing system}

Companies will have to customize their software or hardware to meet the specific requirements of the
customer's system, such as a warehouse management system (WMS). This can involve
significant engineering, integration and testing costs. This topic is addressed in section~\label{deployment}.

\subsection{Need for cloud services}

As mentioned in Section~\ref{spatial_reasoning} if some companies want to deploy generalist humanoid robots, they will have to rely on a robotics transformer/VLM/VLA. Such model cannot run on embedded GPUs and must be deployed on a multi-TPU cloud service. It raises several issues:
\begin{itemize}
	\item The need to develop a dedicated protocol to exchange data between the robot and the VLM;
	\item Quarantee at least a realtime compatible with the tasks assigned to the humanoid;
	\item Due to the protection of data such as the \emph{General Data Protection Regulation (GDPR)}  in EU, it may be impossible to transfer data gathered by humanoids back to a cloud center in USA. Companies may be forced to deploy a data center in different regions of the globe which lead to additional costs that may impact their revenues.
	\item Another burden not addressed in this article will be the need to set up secure communication~\cite{BOTTA2023200237}.
\end{itemize}

\subsection{Communication}

Reliable communication channels will be required:

\begin{itemize}
	
\item By companies that will need a connection to a remote data center either to download new task models in the humanoid or upload data to train a LLM;

\item By companies that will need a internet connection to allow an operator to teleoperate humanoids;

\item By companies to remotely monitor the status of the humanoids;

\item By customers with access to information about their fleets' performance and status which is critical for
customers to understand how their humanoid robots are functioning and to identify any issues; 

\item By customers with access to a customer care front office.

\end{itemize}

Providing customers with a seamless and convenient experience is essential to a successful deployment-
however, integrating various communication channels can be complicated and costly, requiring significant IT
resources and infrastructure at the cost of the companies. And this cost will increase as the fleet of robots (and customers) will increase. 

\subsection{Supervision}

Why remote supervision? Because humanoids are supposed to be autonomous 24/7 but they can get stuck for different reasons. However, if the company promised a 24/7 service,  and the robots are autonomous let us say 60\% of the working time, the
company needs to set up a remote supervision – if agreed with the client - to comply with contractual commitments. 

If humanoids offer a high availability and are equipped with an
auto-diagnostic solution, it can greatly mitigate the need of an active remote
supervision.

Hire staff to remotely monitor humanoids is related to the degree of
autonomy/robustness of the deployed solution. The less robust the
humanoids are, the more staff you need.

This aspect has to be quantified at the early stage of the project because the
burden of the cost induced by the supervision has to be included in the
associated financial offer.

Remote supervision involves:
\begin{itemize}
    \item R\&D team has to develop and maintain supervision tools or purchase one in last resort ;
    \item Dedicated Hardware (servers) and Ethernet connection – probably
    redundant setup will be required. Another option is to go for a cloud
    solution if the bandwidth/response time requirements are compliant with the
    SLA agreed by the client;
    \item Hire and train a team to supervise the different sites.
\end{itemize}
Remote supervision is an important indicator of the robustness/availability of a company solution :
\begin{itemize}
\item Frequency of opened tickets provide an indication about the maturity of an “on site solution”;
\item Frequency of the interventions, i.e. help a robot which is stuck in an
undesirable situation, provide an indication about a use case flaw;
\item If the team grows significantly with the number of sites, it means that
there are (basic) issues which need to be solved still. 
\end{itemize}


\section{Can we outlook the best practices to deploy humanoid robots on industrial sites?\label{deployment}}


\subsection{Pilot: milestones and challenges}

The deployment of new robotic technology can be decomposed into several stages:

\begin{itemize}
	
\item There is a first step which consists of evaluating whether this technology is sufficiently mature to be used in an industrial environment.

To do this, the customer in contact with the robot supplier identifies a very specific task. A simple task like transporting an object from point A to point B or sorting objects for example. The humanoid is placed in a secure perimeter to avoid accidents with customer staff. Another important benefit of a secure perimeter is to ensure that the environment does not change.

This makes it possible to evaluate repeatability, the rate of failure to complete the task, humanoid failures, the number of times the humanoid falls (which is a severe incident in terms of safety), the number of manual or teleoperation handovers when the humanoid is stuck, for example. The goal is to measure the industrial maturity or reliability\footnote{Reliability is the probability of a machine operating without failure.} of the humanoid. In this phase, there are often hardware and or software modifications that are made to the humanoid to correct bugs and improve its robustness. Meantime, the company gather a lot of information  for evaluating maintenance costs.

The evaluation is also critical for both parties to evaluate what we often call "hidden costs":
\begin{itemize}
\item Time to modify or adapt customer infrastructure;
\item Network dependence, integration with other physical and software customer systems;
\item Time to configure and/or train the robot;
\item Time to train employees to setup and use the humanoid;
\item Eventually time spent creating and editing 3D maps for humanoid localization and navigation.
\end{itemize}

Finally measure the acceptance of the robot by the workforce is also a critical factor. If the humanoid is not intuitive and user-friendly, they will be frustrated as they spend countless hours figuring it out or trying to get help from the company.

This stage is an opportunity for the company to improve the humanoid setup, humanoid operation in its environment and maintenance directives.

\item If the evaluation is satisfactory, the customer identifies whether there is a gain or not and which one:
\begin{itemize}
\item Productivity gain: the humanoid costs less than an employee and it works faster and/or longer;
\item Health and social gain: the humanoid carries out tiring or very repetitive and unrewarding tasks, which results in fewer accidents, less sick leave or a reduction in turnover.
\end{itemize}

\item If the customer considers that the gain is significant, then the deployment phase can begin. It is very likely that the humanoids will operate in secure environments initially to preserve the safety of employees, the time necessary to certify that the humanoid robots are in compliance with the safety standards of the industry in question.

\item The last step in deployment consists in taking the humanoid out of the secured perimeter and putting it to work in the factory with other human employees. This step is very complex because the humanoid must be able to plan a sequence of tasks, including navigation, while facing a dynamic and noisy environment, it must pay attention to workplace signs, stay inside walkways, stay alert for vehicle traffic (forklifts, order pickers...) and so on. On the other hand, the humanoid robots must be safe for surrounding workers and compliant with safety standards which has not been defined yet - see section~\ref{safety} for more details.I do not think that this step is reachable in a near future.

\end{itemize}


Following these steps is very important. Many robotics startups fail because they deployed robots on many customer sites without having completely tested their robots. Obviously, the robots did not work as expected, environmental constraints which were not anticipated, faulty mechanics. They were then necessary to send personnel on sites, carry out mechanical repairs and/or changes of parts on the robots. If this happens repeatedly across multiple sites, the cost of maintenance can quickly become uncontrollable and put the startup in a dangerous financial situation without counting on the degradation of the startup's brand image. 

\subsection{Robot robustness and robot supplier pragmatism : two examples}

When a company wants to address a market quickly, a strategy consists of adapting its product to the constraints imposed by the environment and targeted use cases, which sometimes means that it is necessary to make choices about the design of the robot (which can also be considered as sacrifices for researchers). For instance, if the robot has to handle boxes of those shapes and sizes, are unique and known, rather than designing a universal hand that can theoretically take any object, but with a non controllable probability of success, it is wiser to design a gripper adapted to the use case that will work at least 95\% of the time. This point leads us to the robot reliability. Indeed a pilot will have to pass the different steps described above and reliability is a key asset for the company.

\subsubsection{Agility Robotics}
Let's take the case of Agility Robotics which, in my opinion, has a good approach to the market.

In October 24th 2023, Agility Robotics announced that Amazon will begin testing Digit for use in their
operations~\cite{agility_robotics_amazon}.

\begin{figure}[!h]
	\centering
	\includegraphics[scale=0.4]{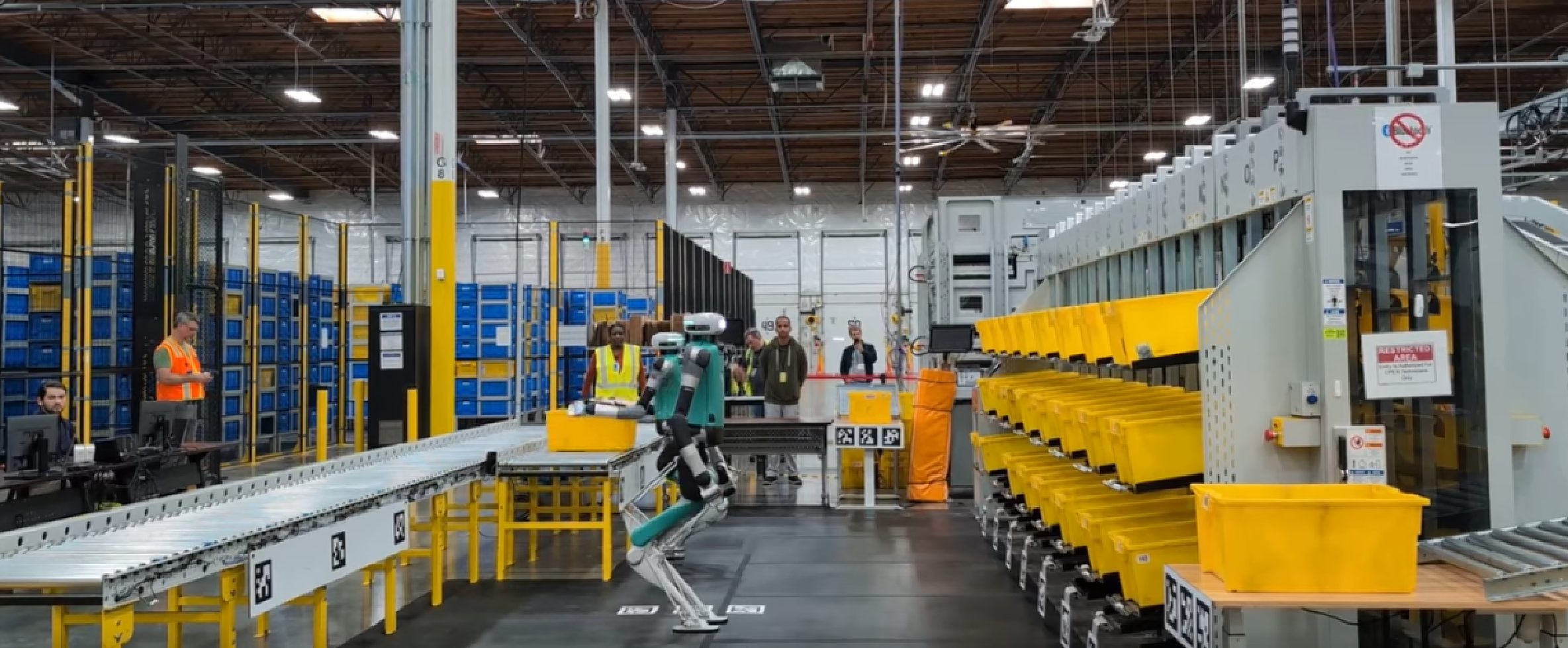}
	\caption{Digit to Assist Amazon Employees with Repetitive Tasks - Credit: Amazon.}
	\label{fig:agility_amazon}
\end{figure}

Digit will first be tested at Amazon’s robotics research and development
facility just south of Seattle. Amazon’s initial use for Digit is to recycle empty totes. It takes totes off a
stationary storage rack and carries them to a conveyor. From available images~\ref{fig:agility_amazon}, we can make the following remarks:

\begin{itemize}
\item The humanoid operates in a secure perimeter closed to Amazon employees; 

\item One can notice that there are "April tags" probably on one hand to localize the robots and on the other hand to label the space (in front of shelves, different positions along the conveyor...); 

\item Humanoids operate in a space clear off obstacles.The environment is structured so that the robots will be able to carry out the task without contingencies; 

\item The task is probably event based, a tote is pushed in an inclined mobile shelve, a sensor may detect the presence of the tote which triggers a digit robot to move in front of the shelve, grasp the tote and carry it to the conveyor belt.
\end{itemize}

In December, Agility Robotics announced a new pilot with GXO Logistics~\cite{robotreport_agility_gxo} at Spanx’s facility in
Flowery Branch, Ga. In the pilot, Digit is moving totes off of autonomous mobile robots (AMRs) and onto a conveyor. The AMRs brings the packed totes to a transfer station. Digit uses its perception system to detect the AMR has arrived. Next, Digit picks up a tote off the top or bottom shelf, carries it over, and places it onto the conveyor. In the future, Agility said Digit will also communicate directly with the AMR fleet manager.

\begin{figure}[!h]
	\centering
		\includegraphics[scale=0.4]{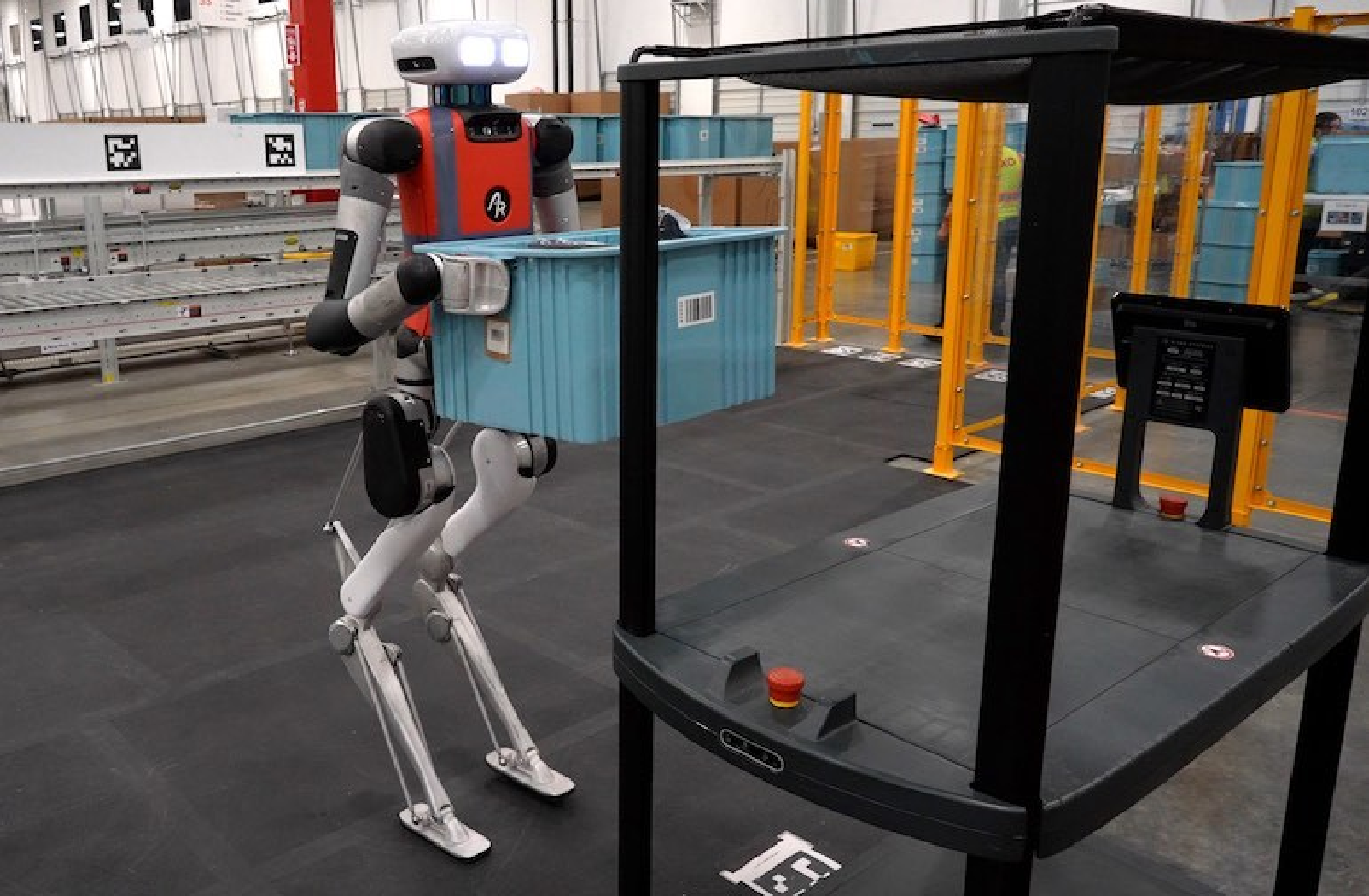}
	\caption{Digit is moving full totes from mobile robots to a conveyor. Credit: Agility Robotics.}
	\label{fig:agility_gxo}
\end{figure}

From available images~\ref{fig:agility_gxo}, we can make the following remarks:
\begin{itemize}
	
\item The robots operate in a secure perimeter closed to Amazon employees; 

\item The task is slightly more complex than the Amazon pilot, in the sense that there is an {AMR} that carries totes filled with products. Once the ARM reached it destination, a message is probably sent through the back office to the Digit. In this case Digit will carry "heavy" totes from the AMR to the conveyor. It will be interesting for Agility Robotics to observe how the actuators of both  arms and shoulders will behave over a long run as well as the ability of the dynamic walking gait to compensate for the additional weight;

\item One can notice that there are "April tags" again so the same algorithms are probably reused;

\item Robots operate in a space clear off obstacles.The environment is structured so that the robots will be able to carry out the task without contingencies. 
\end{itemize}

The tasks are quite simple and rely on a set of robust and proven algorithms such as: digit walking gait has been developed, improved and tested for a decade, perception may be achieved by a deepnet trained to recognize and localize totes, the 3D environment has probably been scanned and created by a SLAM algorithm, efficient localization algorithms are available to localize the robot and may be re-initialized with the april tags, grasping with a dedicated gripper has been intensively tested before going for the pilot. 

\subsubsection{Enchanted Tools}

Enchanted Tools is French startup that’s targeting application spaces like hospitals, nursing homes, and medical clinics with its Mirokai robot, a humanoid torso mounted on a single spherical wheel. The cobot takes an interesting approach with its animated face built around a compact rear projection system. It will interact directly with people and can carry small loads, around 3kg, using a pair of arms.

\begin{figure}[!h]
	\centering
	\includegraphics[scale=0.4]{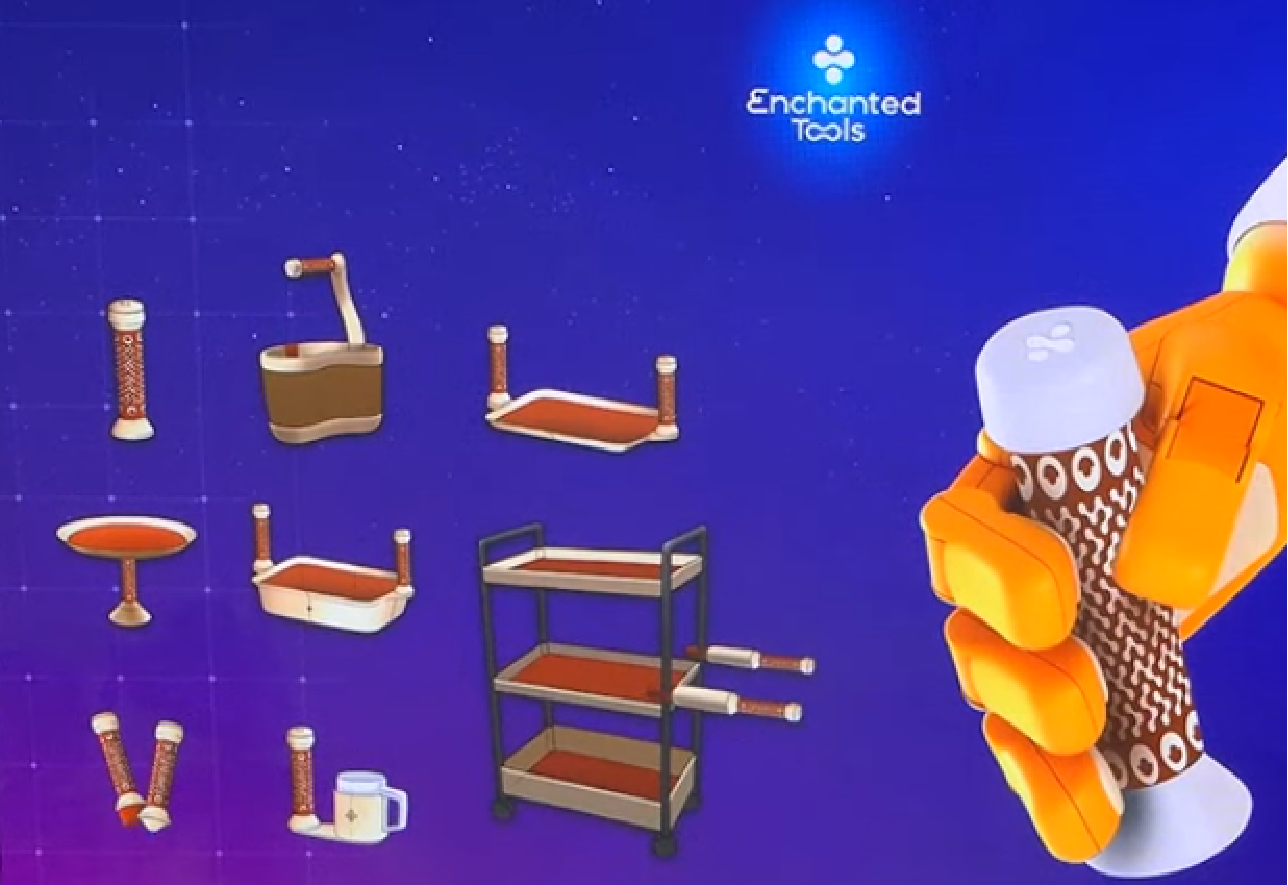}
	\caption{To be written. Credit: Enchanted Tools.}
	\label{fig:enchanted}
\end{figure}

Enchanted Tools offers a pragmatic solution for grasping objects. The robot can only grasp handles with a specific shape. These handles, as can be seen in the figure~\ref{fig:enchanted}, are attached to various objects such as trays or trolleys which will be used by the robot in  applications for which it was designed. The robot has been trained to recognize and grasp these handles only.

Enchanted tools has also developed the concept of "rune". It is a connected object which communicates via Bluetooth with the robot. The rune can also be located by the robot and can be programmed to be labeled as an object such as vase, the tray number xxx, a place like bedroom-xxx or a kitchen for instance.

These runes can be used in the following way:
\begin{itemize}
	
\item  A rune is placed in a room with the label room-xxx and one can ask the robot to move a tray in room-xxx. The robot locates the rune with the label room-xxx and plans a path to the rune's position;
\item A rune can be fixed on a handle with the label tray-yyy, and this handle is fixed on a tray. One can ask the robot to pick up the tray-yyy aand bring it to room-yyy.
\end{itemize}

This robot is being tested in Parisian hospitals to help hospital staff to move stuff around. With this device, the deployment of the robot is faster and above all more efficient. 

\subsection{Conclusion}
My claim here is when a robot supplier wants to sell its robot and bring it to industrial environment and complain with productivity constraints, it must rely on mature technologies, it might not be state-of-the-art, but these technologies have been thoroughly tested so that you can trust them. The robot supplier must show with the pilot(s) that its robot is capable of achieving the task at least 95\% of the working time and this is very hard to accomplish with such complex machinery. This is not a proof of concept in a laboratory for pursuing grants. This does not prevent the company and its R\&D team from developing new algorithms or exploring possible applications of generative AI for instance and maintain a roadmap to incomporate them in the future.


\section{Conclusion}

In the space of a year, a large number of companies specializing in humanoid robots showed up on the international scene. This is an unprecedented situation because these robots are complex, requiring expertise ranging from mechanics to artificial intelligence, and the objective of these companies is to commercialize these robots by 2025!

We are witnessing strategic movements in the high-tech industry. For example, OpenAI is betting on humanoid robots ~\cite{businessinsider_2024} as a source of profit in the medium term. OpenAI is partnering with Figure to equip Figure 01 with high-level visual  based situation analysis capabilities, task planning and language intelligence and is also a shareholder in 1X a competitor of Figure. NVIDIA has recently announced Project GR00T (Generalist Robot 00 Technology), a general-purpose foundation model for humanoid robots, acts as the mind of robots, making them capable of learning skills to solve a variety of tasks~\cite{IEEEspectrum_nvidia}.

In this article we studied 12 companies. We also reviewed the technical challenges posed by these machines, the operation and support aspects once they are deployed and detailed current pilots which tell us about the maturity of these humanoids and the companies' strategy. 

There are companies that want to conquer market share at a fast pace and this is why their humanoid robots rely on proven technologies, the reliability of which is currently tested with pilots.

Other companies choose to rely on innovative technologies such as generative AI, still at the research stage. These technologies theoretically make it possible to address markets other than logistics such as healthcare, household service and/or companionship, but they have not proven their reliability or robustness yet. In addition, these technologies require complex infrastructures to operate. They will be more expensive to implement and maintain. It is a bet on the future, risky by definition.
 
If we take the analogy with self-driving cars, many of these companies promise a general purposes humanoid robot for 2025 or 2026 but it is very likely, as for autonomous cars, that the promises will not be kept and we will have to wait 5 or 10 years to see this type of robot actually operational. However, it is likely that these robots will work in secure perimeters for a few years, while regulation entities define security standards for humanoid robots.

The next few years will be really exciting to follow, there are already pilots going on and we are seeing some humanoids at work in warehouses in secure perimeters. Some pilots will be successfull and trigger the "mass" production of these humanoids. While other pilots will highlight the current limitations of these prototypes, which will initiate new research  to resolve them. It will lead to more agile, reliable and efficient humanoid robots in a near future.

\newpage
\tableofcontents
\newpage
\bibliographystyle{plain}
\bibliography{humanoid}

\newpage
\appendix 
\section{Short introduction of the competitors\label{menagerie}}

\subsection{Appolo A1 from Apptronik}

\begin{table}[ht!]
	\begin{center}
		\begin{tabular}{ p{15cm} }
			
			\toprule
			Appolo A1 From  \href{https://apptronik.com/product-page}{Apptronik} \hfill  \includegraphics[scale=0.1]{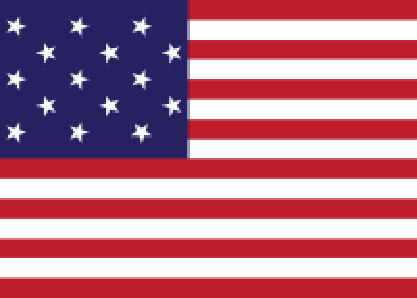}\\ 
			\cmidrule(lr){1-1}
			
			\begin{wrapfigure}{l}{0.25\textwidth}
				\centering
				\includegraphics[scale=0.5]{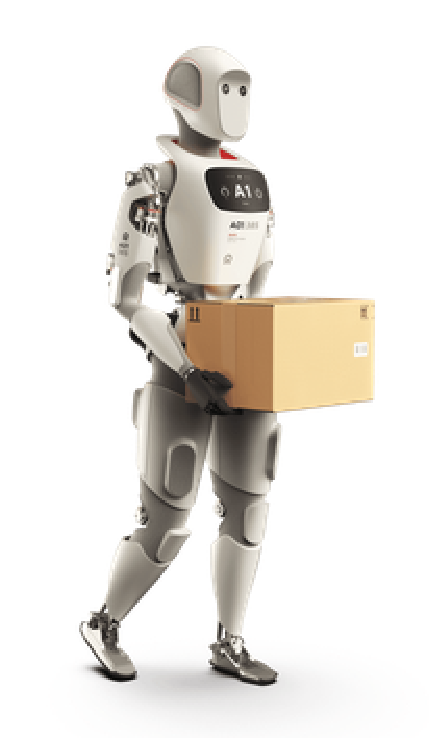}
			\end{wrapfigure}
			
			\textbf{Robotics background} (1) : The founders of Apptronik were heavily involved in making Valkyrie, NASA's first-ever bipedal robot that was initially conceived for the 2013 DARPA Robotics
			Challenge, while they were still working in the lab at the University of Texas at
			Austin. Actually Apptronik was created in 2016 to commercialize the work out of NASA.
			Since then, the company built 13 robots.
			
			\textbf{Teleoperation} (1): Apptronick developed a VR teleoperation approach for their upper-body humanoid Astra~\cite{IROS22_Jorgensen}. Apptronick short to medium term approach might be hybrid autonomy, with a human overseeing first a few and eventually a lot of Apollos with the ability to step in and provide direct guidance through teleoperation when necessary~\cite{IEEEspectrum_apptronik}. 
			
			\textbf{Modularity} (1) modular: Apollo is modular at the torso. So if you want to put
			it on wheels, you can put the upper body on wheels. It’s modular at the end effectors, it’s also modular
			at the head as well in terms of putting different sensor payloads on it.
			
			\textbf{Dexterous hands} (0.6) Apptronik released a \href{https://www.youtube.com/watch?v=SN3GQQCTY4Y}{video} showing Appolo A1 manipulating a tote.
			
			\textbf{Task planning} (0.2): The topic has been addressed~\cite{johnkoetsier_apptronik} but I did not see how Apptronik is tackling these problem.
			
			\textbf{AI} (0.4): It is difficult to tell because Apptronik did not mention if grasping is achieved through deep learning or not.
			
			\textbf{Walking gait} (0.6) : Apptronik released a \href{https://www.youtube.com/watch?v=uJOA5IDaL5g}{video} from which we can bet that Appolo A1 walking speed is around 2/3 m/s.
			
			\textbf{Market} (0.5) The early version of the Apollo humanoid is aimed at handling tasks in the
			logistics and manufacturing industries (palletizing, moving totes, item picking - probably interacting with AMRs). In a talk with John Koetsier~\cite{johnkoetsier_apptronik}, Jeff Cardenas, the {CEO} of Apptronik, mentioned pilots and we can expect some PR in 2024.
			
			\\ \bottomrule
			\begin{multicols}{2}
				\begin{itemize}
					\item Height: 173cm
					\item payload: 25kg
					\item weight: 73kg
					\item runtime: 4hours swappable battery
					\item walking speed: 1 m/s.
					\item Degrees of Freedom: 41 (hand 6DoF)
					\item Perception: Depth cameras
					\item target price: US\$ 50k
				\end{itemize}
			\end{multicols}
			\\ \bottomrule
		\end{tabular}
		\caption{Short introduction of Appolo A1 from Apptronik}
		\label{apptronik_table}
		
	\end{center}
\end{table}

\newpage

\subsection{Phoenix from Sanctuary AI}

\begin{table}[ht!]
	\begin{center}
		\begin{tabular}{ p{15cm} }	
			
			\toprule
			Phoenix the general-purpose hmanoid robot from \href{https://sanctuary.ai}{Sanctuary AI}  \hfill  \includegraphics[scale=0.15]{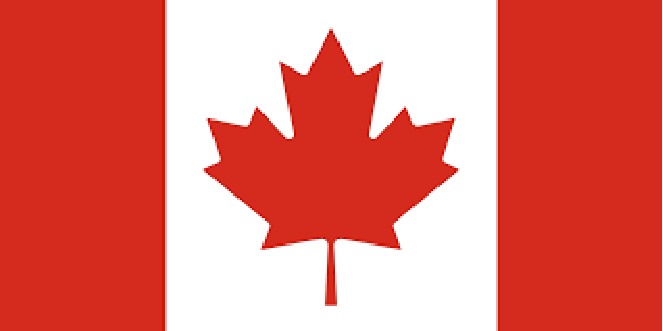}\\ 
			\cmidrule(lr){1-1}
			\begin{wrapfigure}{l}{0.25\textwidth}
				\centering
				\includegraphics[scale=0.5]{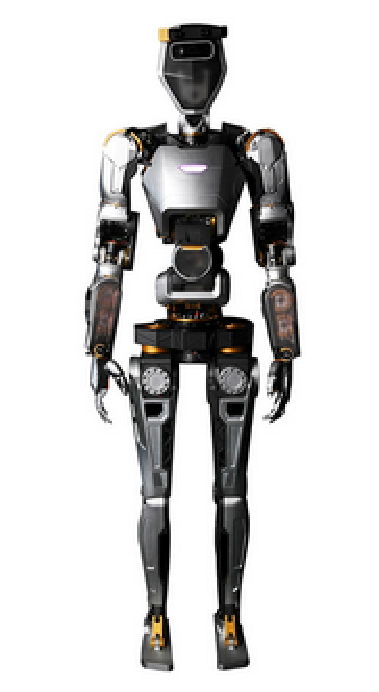}
			\end{wrapfigure}

			\textbf{Robotics background} (0.5): Sanctuary {AI} has been founded in 2018 by Geordie Rose, Suzanne Gildert and Olivia Norton. All of them having top level position in Kindred {AI}, a company specialized in {AI}-powered manipulator robots which has been sold to Ocado in 2020. The founders are experienced in finance, management and in reinforcement learning.
			
			\textbf{Teleoperation} (1): Sanctuary {AI} considers teleoperation such as directly piloted (or teleoperated) if there is a task  the robot cannot perform autonomously, the pilot can take control of the robot to complete the work while simultaneously helping to further train the robot (imitation learning). Actually high-quality teleoperation is used to collect lot of data about complex or high value tasks (see Market bullet point below) that will be used to feed a cognitive architecture~\cite{IEEEspectrum_phoenix}. 
			
			\textbf{Modularity} (1):  Phoenix right now is either static and tethered, which means it does not move or  on a four wheel base when limited untethered movement is required. A bipedal robot is not their top priority. 
			
			\textbf{Dexterous hand} (1): "Sanctuary {AI} is in some ways a hand delivery mechanism" to quote his CEO~\cite{johnkoetsier_sanctuary}. As Phoenix is willing to design a  humanoid robot that is \textbf{truly general-purpose}, a dexterous human-like hands is mandatory. This is why Sanctuary {AI} is actively working on  visual servoing, real-time simulation of the grasping process, and mapping between visual and haptic data.
			
			\textbf{Task planning} (0.2): what Sanctuary {AI} wants to achieve is "you speak to the robot where you issue it a command and the robot has to interpret what you mean. And then in the context in which it is in the world, execute that command for you"~\cite{johnkoetsier_sanctuary}.To do so Sanctuary {AI} claims that the control system (called Carbon) “enables Phoenix to think and act to complete tasks like a person".
			
			\textbf{AI} (0.6): From what I can read, I can assume that Sanctuary {AI} aims at developing  End-to-end Perception-grasping deep network either though reinforcement learning or other methods.
			
			\textbf{walking gait} (0): as mentioned above, a bipedal robot is not in their priority although there are planning to work on it.
			
			\textbf{Market/Pilot} (0.5): Thanks to the unique hand dexterity of Phoenix, Sanstuary AI is going to focus more on higher value tasks, the ones which are very difficult or expensive to do. This is why Phoenix is looking very carefully at fly in, fly out jobs. Phoenix has already signed pilots through its partnership with Canadian Tire Corporation (CTC). 
			
			\\ \bottomrule
			\begin{multicols}{2}
				\begin{itemize}
					\item Height: 170cm
					\item payload: max 25kg
					\item weight 70kg
					\item runtime: 5hours
					\item walking speed: max 1.5 m/s
					\item Degrees of Freedom for the hand 20 DoF
					\item Perception: depth cameras
					\item target price: Not disclosed
				\end{itemize}
			\end{multicols}
			\\ \bottomrule
		\end{tabular}
		\caption{Short introduction of Phoenix from Sanctuary AI}
		\label{phoenix_table}
		
	\end{center}
\end{table}

\newpage

\subsection{Figure 01 from Figure}

\begin{table}[ht!]
	\begin{center}
		\begin{tabular}{  p{15cm} }
			\toprule
			Figure 01 from \href{https://figure.ai}{Figure} \hfill  \includegraphics[scale=0.1]{usa.eps} \\ 
			
			\cmidrule(lr){1-1}
			
			\begin{wrapfigure}{l}{0.25\textwidth}
				\centering
				\includegraphics[scale=0.4]{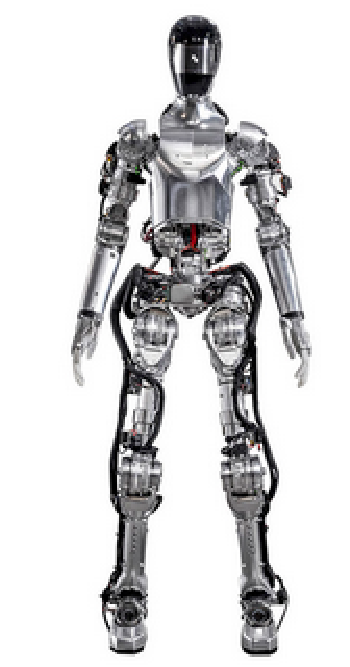}
			\end{wrapfigure}
			
			\textbf{Robotics background} (0.5): The CTO of Figure is Jerry Pratt~\cite{IEEEspectrum_Figure}, a former PhD student of Gill Pratt (the man behind the DARPA Robotics Challenge)  who did his thesis on Flamingo~\cite{JPratt_PhDthesis}. Jerry Pratt spent 20 years at the Florida Institute for Human and Machine Cognition (IHMC), where he led the team that took second place at the DARPA Robotics Challenge Finals. He used to work with DRC Atlas, NASA’s Valkyrie, and most recently Nadia. Jerry Pratt is an important asset to Figure because he knows very well how to make  humanoids walking. Not only he has spent his entire career developing algorithms, actuators and hardware for humanoids but he is experienced and he gained a deep understanding of dynamic walking.
			
			\textbf{Teleoperation} (0.5): Figure relies on different technologies to train their humanoids. Tele-operating to the robot and instructing them on how to pick up a bin, box, or an object on the table is one of them~\cite{newatlas_Figure}.
			
			\textbf{Modularity} (0): Figure is planning to sell a full humanoid robot.
			
			\textbf{Dexterous hand} (0.6): from this \href{https://www.youtube.com/watch?v=Q5MKo7Idsok}{video} one can see the dexterous capabilities of Figure with its 6 DoF hands~\cite{robotReport_figure01}.
			
			\textbf{Task planning} (0.7):Figure is partnering with OpenAI to demonstrate in this \href{https://www.youtube.com/watch?v=Sq1QZB5baNw}{video} that the Figure 01 is able to understand verbal order and plan tasks.  
			
			\textbf{AI} (0.6):Although the dynamic walking is probably achieved with "newtonian" control and optimization algorithms, Figure is relying on an AI team which is running humanoid robots with end-to-end neural networks performing highly complicated and dexterous tasks ("end-to-end" means from perception to grasping).  The big picture is to teach robots how to do tasks, and as the robot fleet grows so will the training sets. Tasks and/or skills learned by a group of robots in a factory are transferred to the other member of the fleet. 
			
			Figure claimed in this  \href{https://www.youtube.com/watch?v=Q5MKo7Idsok}{video} that it took 10 hours watching videos to train their humanoid to prepare a Coffee. It looks like that the learning strategy followed by Figure is watching videos. This is a well done marketing video. We have to take it for granted. At least what we observe is how fluid are the movements of arms, hands and fingers.
			
			\textbf{walking gait}(0.6):Figure.ai released a \href{https://www.youtube.com/watch?v=jACJruCzUzY}{video} from which we can bet that Figure walking speed is around 2/3 m/s.
			
			\textbf{Market/Pilot} (0.5): Figure~\cite{reuters_figure_bmw} announced a commercial agreement with BMW to supply the automaker with humanoid workers for their manufacturing plant in South Carolina. This is a good news for the company and I am eager to see what kind of tasks will be assigned to Figure 01 and how it will perform.
			
			\\ \bottomrule
			\begin{multicols}{2}
				\begin{itemize}
					\item Height: 168cm
					\item payload: 20kg
					\item weight 60kg
					\item runtime: 5hours
					\item walking speed: 1.2m/s
					\item Degrees of Freedom: 41 (hand 6 DoF)
					\item Perception: Not disclosed
					\item target price: Not disclosed
				\end{itemize}
			\end{multicols}
			\\ \bottomrule
			
		\end{tabular}
		\caption{Short introduction of Figure 01 from Figure AI }
		\label{figure_table}
	\end{center}
\end{table}


\newpage

\subsection{RAISE-A1 from Agibot}

\begin{table}[ht!]
	\begin{center}
		\begin{tabular}{  p{15cm} }
			\toprule
			RAISE-A1 general purpose robot  from \href{https://agibot.com}{Agibot} \hfill  \includegraphics[scale=0.1]{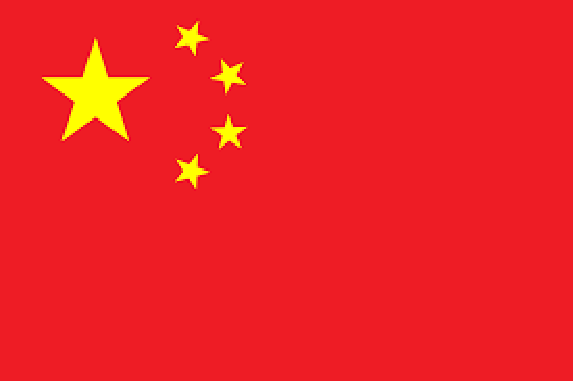} \\ 
			
			\cmidrule(lr){1-1}
			
			\begin{wrapfigure}{l}{0.25\textwidth}
				\centering
				\includegraphics[scale=0.3]{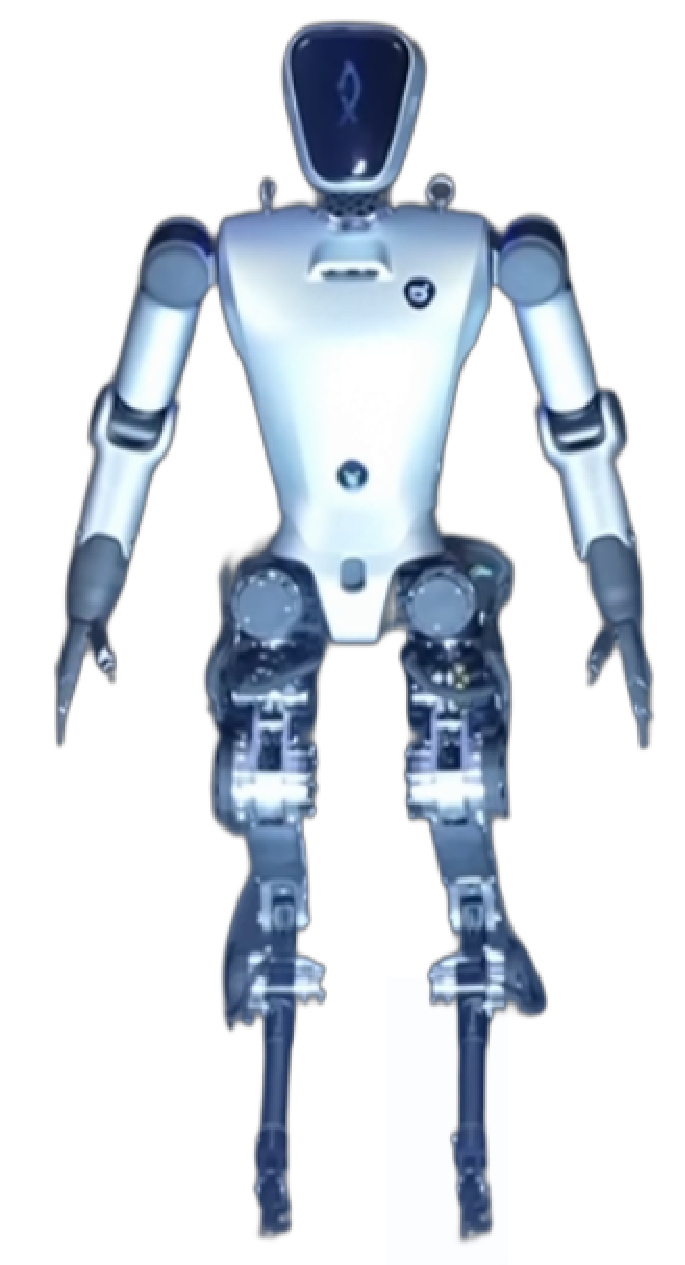}
			\end{wrapfigure}
			
			\textbf{Robotics background}(0.5): Peng Zhihui the founder of Agibot joined Huawei in November 2020 through the Chinese tech giant’s so-called “Geniuses” recruitment program as an AI algorithm engineer. In 2022 he left Huawei to create Agibot. In the \href{https://www.youtube.com/watch?v=ZwjxbDVbGpU}{2023 Agibot Humanoid Robot RAISE-A1 Launch Event} Peng presented a quite detailed description of A1 as well as his vision for the coming years. As you can see from the picture A1 is very similar to Digit. Agibot did not promote A1 outside of China.
			
			\textbf{Teleoperation} (0): there is no mention of teleoperation in the Agibot presentation.
			
			\textbf{Modularity} (1) : Agibot is proposing the upper body of ARSE-A1 in a static and tethered version on wheels as well.
			
			\textbf{Dexterous hand} (0.6) : the \href{ https://www.youtube.com/watch?v=PIYJtZmzs70}{vide} is showing RAISE-A1 manipuling objects - although the movements are not that fluid.
			
			\textbf{Task planning} (0.6): Agibot has designed an AI architecture which relies heavily on Large Language Models (LLM). They are working on a LLM named "WorkGPT" a large scale data pre-trained language and image model with strong capabilities in semantic understanding, logical reasoning, image recognition and code generation.  The objective is an end-to-end natural language interface to give order to the robot which is able to transform into a chain of tasks.
			
			They divide the robot's cognitive system into four components : "cloud deployment hyperbrain" (WorkGPT which centralizes all the knowledge gathered by the robots fleet and is able to slove new problems) - brain (a light Visual Language Model or VML) - cerebellum (command-level or motion control such as MPC WBC) and brainstem (servo-level).
			
			\textbf{AI} (0.3) : As mentioned above, RAISE-A1 will be equipped with a light VML so I can say at least that the robot will be able to recognize and classify objects. However I cannot tell whether there is an end-to-end deep net from perception to grasping.
			
			\textbf{walking gait} (1): Agibot in the \href{https://www.youtube.com/watch?v=ZwjxbDVbGpU}{2023 Agibot Humanoid Robot RAISE-A1 Launch Event} showed the different stage of the design and test of the biped platform and how engineers challenged the robustness of the walking gait.
			
			\textbf{Market/Pilot} (0.2): The targeted market is industrial manufacturing. However Agibot aims at developing an ecosystem for  developers to foster new applications.
			
			\\ \bottomrule
			\begin{multicols}{2}
				\begin{itemize}
					\item Height: 175cm
					\item payload: 20kg
					\item weight:  53kg
					\item runtime: Not disclosed
					\item walking speed: 2m/s
					\item Degrees of Freedom: 49 (Hand 12 DoF + 5 passive DoF)
					\item Perception: depth (RGBD) cameras
					\item price: RMB 200K
				\end{itemize}
			\end{multicols}
			\\ \bottomrule
			
		\end{tabular}
		\caption{Short introduction of RAISE-A1 from Agibot}
		\label{agibot_table}
	\end{center}
\end{table}

\newpage

\subsection{Optimus from Tesla}

\begin{table}[ht!]
	\begin{center}
		\begin{tabular}{  p{15cm} }
			\toprule
			Optimus gen2 the general purpose humanoid robot from \href{https://tesla.com/}{Tesla} \hfill  \includegraphics[scale=0.1]{usa.eps} \\ 
			
			\cmidrule(lr){1-1}
			
			\begin{wrapfigure}{l}{0.25\textwidth}
				\centering
				\includegraphics[scale=0.5]{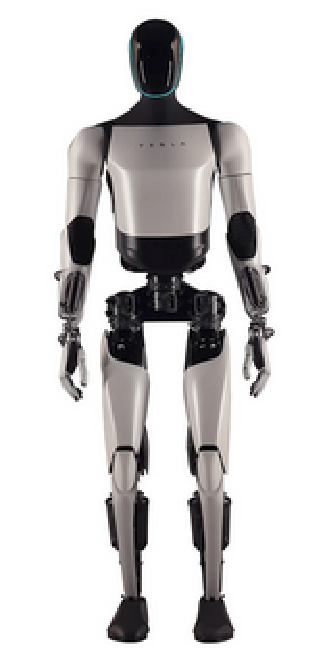}
			\end{wrapfigure}
			
			\textbf{Robotics background} (0.5): Tesla is known for its electric cars and its FSD (Full Self Driving) also called autopilot application which allows its cars to be partially autonomous (L2+ in the jargon of car manufacturers). Tesla was the first company which enters the race with a bipedal robot named Optimus which was unveiled at Tesla’s 2022 AI Day.
			One year later, Optimus gen2  was introduced at Tesla’s 2023 AI Day. Musk said the robot will be able to operate tools and do useful things like carrying and manipulating objects in factories and other settings.  Without knowing it, Tesla triggered a race in which a large number of companies were engulfed.
			
			\textbf{Teleoperation} (0): Tesla used motion capture to demonstrate key frames and then ran them through its optimization program for grasping and balance.
			
			\textbf{Modularity} (0): Optimus has been developed as a full humanoid robot.
			
			\textbf{Dexterous hand} (0.6): Tesla posted a promotional \href{https://www.youtube.com/watch?v=cpraXaw7dyc}{video} of Optimus Gen2 showing its dexterity.
			
			\textbf{Task planning} (0): This topic has not been mentioned at this stage.
			
			\textbf{AI} (0.6) Tesla concentrated it effort on state estimation, control, localization and path planning. However from video we can see that Optimus is equipped with deep net achieving object recognition, classification and segmentation. Elon Musk posted a  \href{https://www.youtube.com/watch?v=8vsTNFUFJEU}{video} on Twitter featuring Tesla's Optimus demonstrating its ability to fold laundry. However, Elon Musk added in a subsequent tweet “Important note: Optimus cannot yet do this autonomously, but certainly will be able to do this fully autonomously and in an arbitrary environment (won’t require a fixed table with box that has only one shirt.
			
			\textbf{Walking gait robustness} (0.6): The \href{https://www.youtube.com/watch?v=cpraXaw7dyc}{video}  which shows Optimus Gen2  walking. Movements are fluid but Tesla never show how robust the walking gait is with respect to disturbances.
			
			\textbf{Market(s)} (0.5): As Tesla is manufacturing cars, Optimus will be able to operate tools and do useful things like carrying and manipulating objects in factories and other settings. Nevertheless Tesla has a significant advantage over its competitors: it has factories and set up pilots to test Optimus gen2 is not an issue. 
			
			\\ \bottomrule
			
			\begin{multicols}{2}
				\begin{itemize}
					\item Height: 173cm
					\item payload: 20kg
					\item weight: 63kg
					\item Battery: 2.3 kilowatt-hour, 52-volt battery pack
					\item walking speed: max 5 m/s
					\item Degrees of Freedom: 50 (hand 11 DoF)
					\item  Perception:  cameras, ultrasonic sensors (same suite as Tesla cars)
					\item target price: US\$ 20K
				\end{itemize}
			\end{multicols}
			\\ \bottomrule
		\end{tabular}
		\caption{Short introduction of Optimus from Tesla}
		\label{tesla_table}
	\end{center}
\end{table}

\newpage

\subsection{H1 from Unitree}

\begin{table}[ht!]
	\begin{center}
		\begin{tabular}{  p{15cm} }
			\toprule
			H1 the general purpose humanoid robot from \href{https://unitree.com/}{Unitree} \hfill  \includegraphics[scale=0.1]{china.eps} \\ 
			
			\cmidrule(lr){1-1}

			\begin{wrapfigure}{l}{0.25\textwidth}
				\centering
				\includegraphics[scale=0.4]{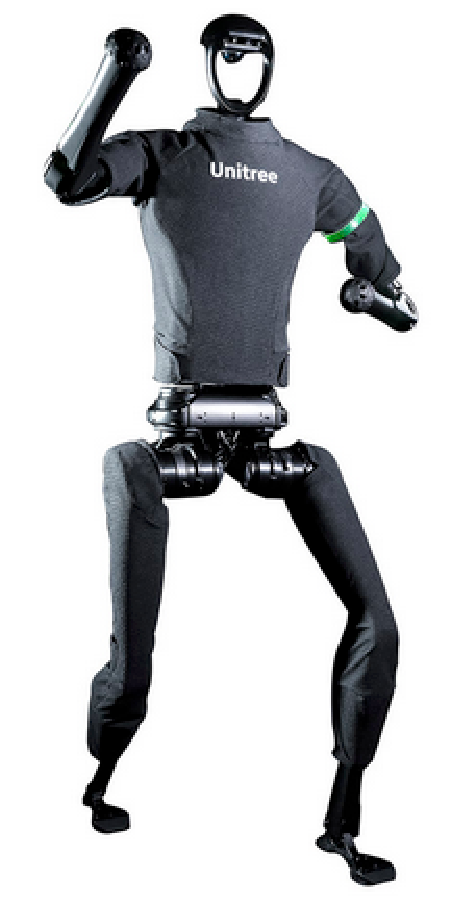}
			\end{wrapfigure}
			
			\textbf{Robotics background} (1): Unitree is a world leader in the design, development and manufacturing of quadruped robots. Unitree designs its own actuators, locomotion and perception algorithms. They have accumulated a lot of experience which allow them to go from a quadruped robot to a biped robot~\cite{robotReport_Unitree} without much effort.
			
			\textbf{Teleoperation} (0): H1 is not teleoperated so far.
			
			\textbf{Modularity} (0): H1 has been developed has a full humanoid robot.
			
			\textbf{Dexterous hand} (0): H1 is not equiped with hands so far.
			
			\textbf{Task planning} (0): This is not a topic addressed by Unitree engineers at this stage.
			
			\textbf{AI} (0.2): Unitree engineers focused on perception and navigation using data from a depth camera and 3D Lidar. However we can probably bet that the algorithms regarding objects recognition and classification deveoped for the quadrupeds are ported on H1.
			
			\textbf{Walking gait robustness} (1): The \href{https://www.youtube.com/watch?v=q8JMX6PGRoI}{video}  which shows the H1 humanoid robot walking in an unstructured environment is impressive. Despite the fact that engineers push it, the robot dynamically stabilizes and continues walking. This is clearly the most robust robot on the list. H1 is equipped with very powerful electrical actuators delivering a torque up to 360 N.m at the hip - with such actuators, H1 may be able to walk/run at a maximum speed of 5m/s.
			
			\textbf{Market(s)} (0.2): H1 is considered as a research platform by Unitree, at least for now. Therefore the market is research labs and/or companies which wants to develop applications. Like Agibot and Fourier Intelligence they will rely on a network of labs and a startup network to develop industrial applications. 
			
			\\ \bottomrule
			
			\begin{multicols}{2}
				\begin{itemize}
					\item Height: 180cm
					\item payload: 20kg
					\item weight: 47kg
					\item Battery: 850 Wh
					\item walking speed: max 3.4 up to 5 m/s
					\item Degrees of Freedom: 18 (no hand)
					\item  Perception:  depth camera (D435i) and 3D Lidar (LIVOX MID360)
					\item target price: US\$ 90K
				\end{itemize}
			\end{multicols}
			\\ \bottomrule
		\end{tabular}
		\caption{Short introduction of H1 from Unitree}
		\label{Unitree_table}
	\end{center}
\end{table}

\newpage

\subsection{GR1 from Fourier Intelligence}

\begin{table}[ht!]
	\begin{center}
		\begin{tabular}{  p{15cm} }
			\toprule
			GR-1 general purpose humanoid robot from \href{https://fourierintelligence.com/}{Fourier Intelligence} \hfill  \includegraphics[scale=0.1]{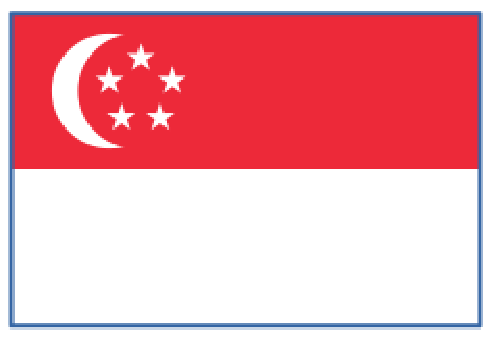} \\ 
			
			\cmidrule(lr){1-1}
			
			\begin{wrapfigure}{l}{0.25\textwidth}
				\centering
				\includegraphics[scale=0.3]{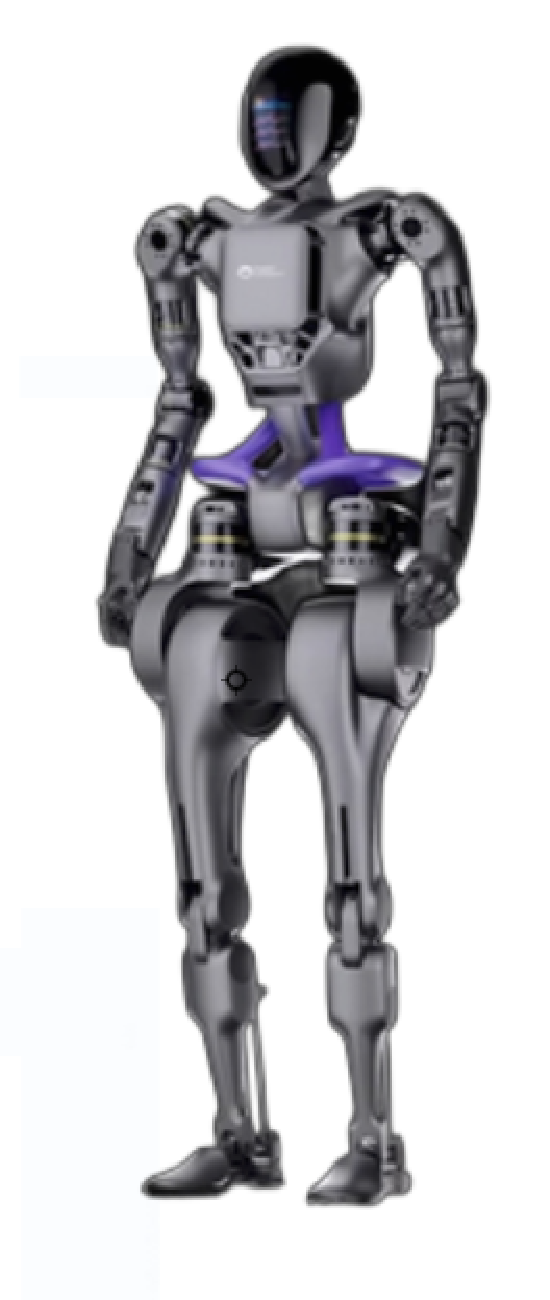}
			\end{wrapfigure}
			
			\textbf{Robotics background}(1) Fourier Intelligence  is developing exoskeleton and rehabilitation robotics since 2015. its \href{https://fourierintelligence.com/rehabhub/}{RehabHub} platform offers a series of integrated physical therapy devices for treating various issues, from wrist strength games to hand and finger grip training, all the way up to lower-body exoskeletons for training people to walk, sit, stand, balance and climb stairs. Their engineers studied the human walking gait and they are familiar with actuators and control algorithms that are necessary to build at least the lower part (from toes to the hip) of a humanoid robot.

			\textbf{Teleoperation} (0): Fourier  Intelligence is not working on the possibility to teleoperate GR1.
			
			\textbf{Modularity} (0) : GR1 is a full humanoid robot - there is no plan to sell the upper body only.
			
			\textbf{Dexterous hand} (0.5): Fourier Intelligence posted videos in which we can GR1 with hands (11 DoF) or a simple gripper.
			
			\textbf{Task planning} (0):  Fourier Intelligence did not mention that it is working on that subject.

			\textbf{AI} (0.2): It looks like that Fourier Intelligence engineers focused on the dynamic walking gait and developed very basic capabilities regarding navigation, perception and grasping. The strategy of Fourier Intelligence consists in producing may be hundred GR-1 and ship them to R\&D labs worldwide where local researchers will develop algorithms to make them more capable~\cite{newAtlas_fourier}.

			\textbf{Walking gait} (1): Fourier Intelligence published \href{https://www.youtube.com/watch?v=xQJkzS0i4w0}{videos} showing an impressive robustness of the robot with respect to engineers pushes and kicks. Notice that the electric motors in the hips, the largest ones, will be capable of generating up to 300 Nm of torque.
			
			\textbf{Market(s)} (0.2): Fourier said it expects to sell the GR-1 for applications such as research and education, concierge and guiding, entertainment and exhibition, industrial production and logistics, healthcare and rehabilitation, safety inspection, household service, and companionship. 
			
			\\ \bottomrule
			
			\begin{multicols}{2}
				\begin{itemize}
					\item Height: 165cm
					\item payload: 20kg
					\item weight: 55kg
					\item runtime: not disclosed
					\item walking speed: max 5 m/s
					\item Degrees of Freedom: 40 (hand 11 DoF)
					\item Perception: depth cameras in head and torso
					\item Target price: not disclosed
				\end{itemize}
			\end{multicols}
			\\ \bottomrule
		\end{tabular}
		\caption{Short introduction of GR-1 from Fourier Intelligence}
		\label{Fourier_table}
	\end{center}
\end{table}

\newpage

\subsection{NEO from 1X}

\begin{table}[ht!]
	\begin{center}
		\begin{tabular}{  p{15cm} }
			\toprule
			Neo from \href{https://www.1x.tech/androids/neo}{1X} \hfill  \includegraphics[scale=0.1]{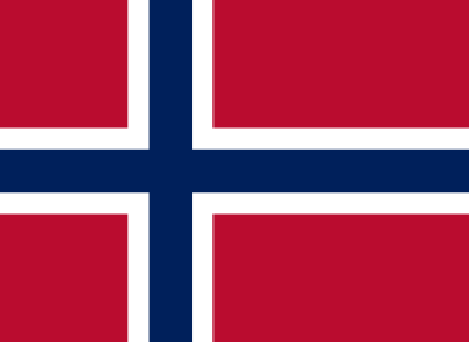} \\ 
			
			\cmidrule(lr){1-1}

			\begin{wrapfigure}{l}{0.25\textwidth}
				\centering
				\includegraphics[scale=0.4]{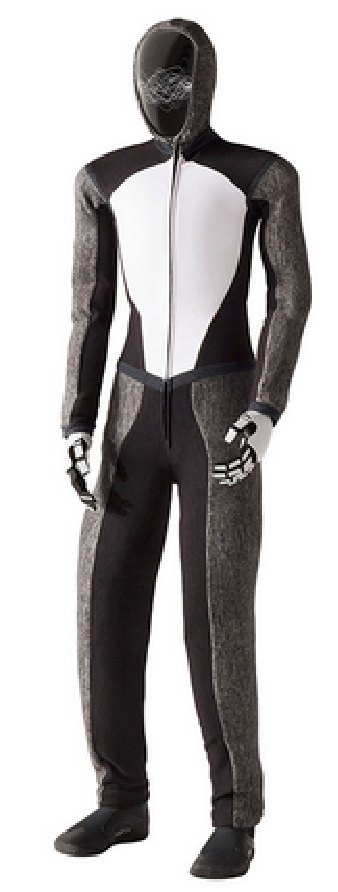}
			\end{wrapfigure}
			
			\textbf{Robotics background} (0.5) : the Norwegian firm closed a funding round of \$23.5 million in April 2023 and {OpenAI} was the round’s lead~\cite{techcrunch_1X_2024}.Given the role {OpenAI} has  played in generative AI’s  rise over the past year. That underlying technology will almost certainly play an outsized role in shaping robotics’ future. OpenAI decided to invest in 1X and will probably work closely with the company. On the other hand, 1X is actively developing {EVE} which is an upper body (torso, two arms and a head) mounted on wheels.
			
			\textbf{Teleoperation} (1) : 1X relies on a team of (tele)operators to train ML models allowing {EVE} to exhibit specific behaviors~\cite{1X_2023_1}. Taking over the humanoid which is stuck, by a teleoperator is also considered.
			
			\textbf{Modularity} (1) :1X is proposing {EVE} which is an articulated upper body on wheels. Although 1X is showing artistic pictures of {NEO}, it is likely that There will be synergies between {EVE} and {NEO}. The AI capabilities can be transfered from one android to the other.
			
			\textbf{Dexterous hand} (0.2): 1X advertizes a  \href{https://www.youtube.com/watch?v=20GHG-R9eFI}{video} in which {EVE} is packing objects in a box with a gripper.

			\textbf{Task Planning} (0.3): As {1X} is backed up by OpenAI, the company is certainly working on that hot topic. However no videos and/or PR have been released so far.
			
			\textbf{AI} (0.6): The recent note published by 1X~\cite{1X_2024_1} shows {EVE} robots handling various tasks with End-to-end Perception-grasping neural nets trained with algorithms based on diffusion models. 
			
			\textbf{Walking gait robustness} (0): 1X is selling {EVE} but there is no videos of {NEO} in action.
			
			\textbf{Market/Pilot} (0): 1X is still working on its humanoid robot and thus it is too soon to mention any pilot. However 1X is willing to address industrial tasks (logistics, manufacturing), support individuals with mobility challenges or helping the robotics community to explore fields like psychology and artificial intelligence.
			
			\\ \bottomrule
			
			\begin{multicols}{2}
				\begin{itemize}
					\item Height: 165cm
					\item payload: 20 kg
					\item weight: 30kg 
					\item runtime between 2 and  4 hours (depending on the payload)
					\item walking speed: 3 m/s
					\item Degrees of freedom: Not disclosed
					\item Perception: Not disclosed
					\item Target price: Not disclosed
				\end{itemize}
			\end{multicols}
			\\ \bottomrule
			
		\end{tabular}
		\caption{Short introduction of {NEO} from 1X}
		\label{1X_table}
	\end{center}
\end{table}

\newpage

\subsection{Digit from Agility Robotics}

\begin{table}[ht!]
	\begin{center}
		\begin{tabular}{  p{15cm} }
			\toprule
			digit from \href{https://agilityrobotics.com/}{Agility Robotics} \hfill  \includegraphics[scale=0.1]{usa.eps} \\ 
			
			\cmidrule(lr){1-1}

			\begin{wrapfigure}{l}{0.25\textwidth}
				\centering
				\includegraphics[scale=0.4]{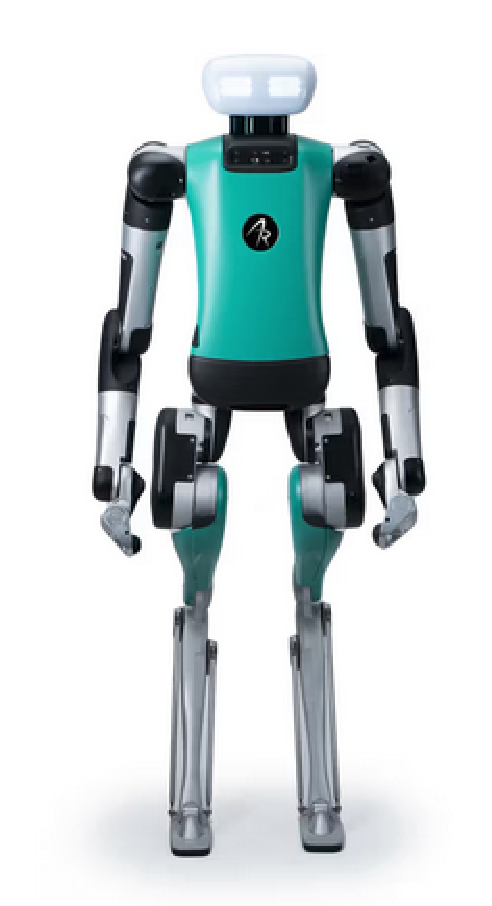}
			\end{wrapfigure}
			
			\textbf{Robotics background} (1) : Digit is the result of a long term research  starting with ATRIAS ~\cite{Ramezani2014PerformanceAA} developed at the Dynamic Robotics lab, Oregon State University (OSU), directed by Jonathan Husrt. It was  followed by Cassie~\cite{Cassie_2023} which was designed by Agility Robotics, the spin off of OSU, those Chief Robot Officer is  Jonathan Husrt. 
			
			\textbf{Teleoperation} (0) : Agility Robotics did not mention if it is using a teleoperation mode.
			
			\textbf{Modularity} (0) : Digit is sold as a full biped.
			
			\textbf{Dexterous hand} (0.2): Digit’s hands are designed to do one thing: move totes (plastic bins that control the flow of goods in a warehouse). They’re not especially humanlike, and do not look fancy, but they’re exactly what Digit needs to do the job that it needs to do ~\cite{IEEEspectrum_agility}. 	
			
			\textbf{Task Planning} (0.7): in a recent \href{https://www.youtube.com/watch?v=CnkM0AecxYA}{video}, Agility Robotics demonstrated that it is possible to give verbal orders to digit and digit was able to plan a set of tasks to execute the order by using a dedicated LLM.  Although the environment is very simple, a set of towers of different heights as well as three boxes, each one with a a different color and pictograms, and the orders like "move the red box one the lowest tower" for instance, digit was able to execute the order. For old researchers, it may recall them STRIPS~\cite{Fikes1971STRIPSAN} the planner created for Shakey by the SRI. Research on using LLM to enable robots to plan a sequence of tasks to execute verbal orders is just beginning, this demonstration showed us that it is possible.
			
			\textbf{AI} (0.4) : Agility Robotics relies on AI for perception. Nevertheless, I cannot tell whether the grasping algorithms relie on C++ or on deepnet.
			
			\textbf{Walking gait robustness} (1) : Digit is the result of at least a decade of research on dynamic walking gaits.
			
			\textbf{Market/Pilot} (1) : Agility Robotics is targeting logistics based tasks. These tasks like moving totes from point A to point B that should be automatized because companies are having a lot of trouble finding people to do them~\cite{IEEEspectrum_agility}. Agility Robotics is very pragmatic in its approach to market, it is running two pilots and we will go into details in section~\ref{deployment}.
			
			\\ \bottomrule
			
			\begin{multicols}{2}
				\begin{itemize}
					\item Height: 175cm
					\item payload: 15kg
					\item weight 65kg 
					\item runtime: not disclosed
					\item walking speed: 1.5m/s
					\item Degrees of freedom:22 
					\item Perception:  (RGBD) depth cameras and a 3D Lidar
					\item Target price: Not disclosed
				\end{itemize}
			\end{multicols}
			\\ \bottomrule
			
		\end{tabular}
		\caption{Short introduction of Digit from Agility Robotics}
		\label{digit_table}
	\end{center}
\end{table}

\newpage

\subsection{XP5 from xpeng}

\begin{table}[ht!]
	\begin{center}
		\begin{tabular}{  p{15cm} }
			\toprule
			XP5 from \href{https://xpeng.com}{Xpeng} \hfill  \includegraphics[scale=0.1]{china.eps} \\ 
			
			\cmidrule(lr){1-1}
			
			\begin{wrapfigure}{l}{0.25\textwidth}
				\centering
				\includegraphics[scale=0.2]{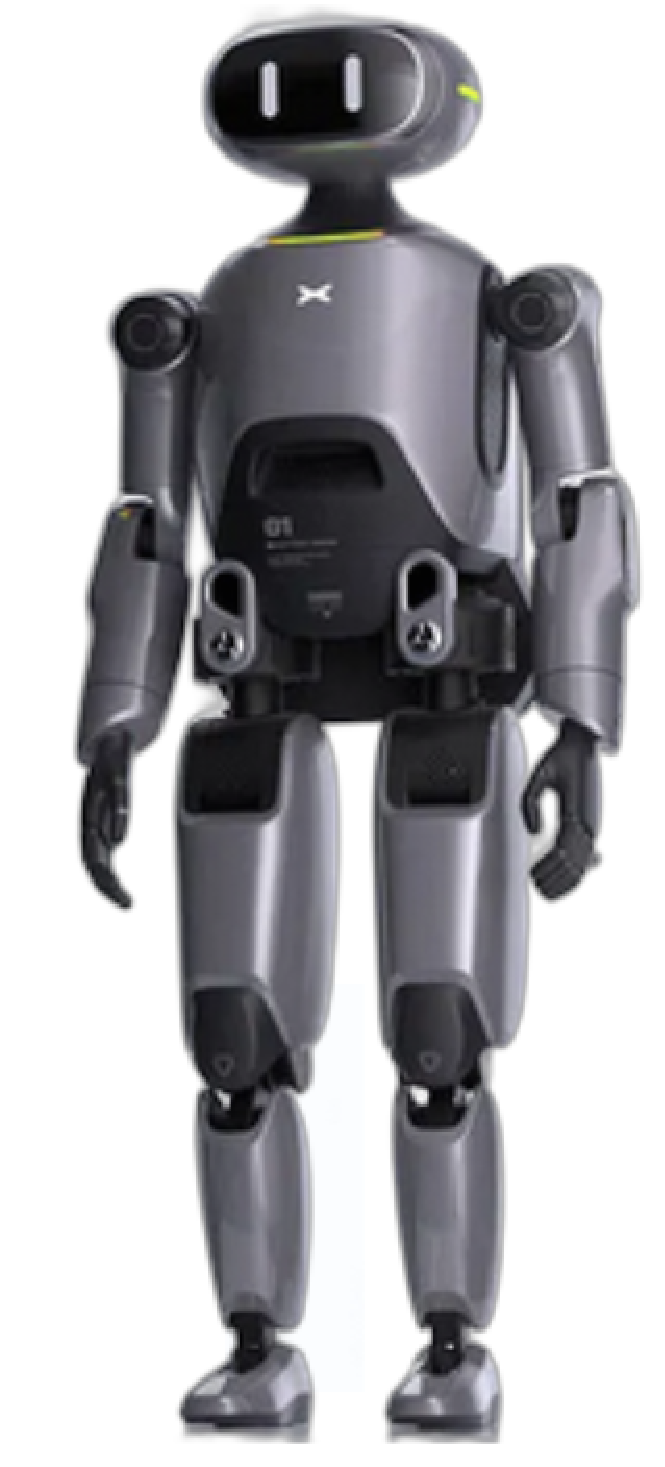}
			\end{wrapfigure}
			
			\textbf{Robotics background} (0.2) Xpeng follows in Tesla's footsteps. Xpeng was established in 2014 and specializes in the development of electric vehicles. Like Tesla, he developed X-NGP advanced driver assistance system and like Tesla every year there is Xpeng Tech Day during which its CEO He Xiaopeng makes announcements on upcoming products.During the  \href{https://www.youtube.com/watch?v=X2dP31FLRwY}{Xpeng Tech Day 2023} the humanoid robotic prototype, the PX5, made its first public appearance~\cite{technode_xpeng}.  
			
			\textbf{Teleoperation} (0): Xpeng did not mention whether it uses teleoperation or not.
			
			\textbf{Modularity} (0): Xpeng is selling a walking humanoid robot only.
			
			\textbf{Dexterous hand} (1) This \href{https://www.youtube.com/watch?v=_q5vFhWiZnA}{video} demonstrated the dexterous robot hand with 11 degrees of freedom which performs tasks like lifting boxes, grasping pens, and pouring water into cups. The robotic hand has 11 degrees of freedom, with two fingers weighing only 430 grams but capable of gripping up to 1 kilogram of load. The mechanical arm has 7 degrees of freedom, a positioning accuracy of 0.05 millimeters, a maximum load of 3 kilograms, and weighs 5 kilograms.
			
			\textbf{Task planning} (0): Xpeng did not mention whether they are investigating task planning or not.
			
			\textbf{AI} (0.4):Xpeng Robotics relies on AI for perception. Nevertheless, I cannot tell whether the grasping algorithms relie on C++ or on deepnet.
			
			\textbf{Walking gait robustness} (1): This \href{https://www.youtube.com/watch?v=BNSZ8Fwcd20}{video}  showed the PX5 humanoid robot which can withstand impact and keep steady even when kicked. Through self-developed high-performance joints, the robot has achieved high-stability walking capabilities and can complete indoor and outdoor walking and obstacle crossing for more than 2 hours.
			
			\textbf{Market/Pilot} (0): According to He, Xpeng is aiming to introduce its PX5 robots to factories and stores by next year’s Tech Day event, utilizing them for tasks such as factory patrolling and in-store product sales.
			
			\\ \bottomrule
			
			\begin{multicols}{2}
				\begin{itemize}
					\item Height: 150cm
					\item payload: unknown
					\item weight: Not disclosed
					\item runtime: 2hours
					\item walking speed: 1-2m/s
					\item Degrees of Freedom: 51 (Hand 11 DoF) 
					\item Perception: Depth Cameras
					\item Target Price: Not disclosed
				\end{itemize}
			\end{multicols}
			\\ \bottomrule
			
		\end{tabular}
		\caption{Short introduction of XP5 from Xpeng}
		\label{xpeng_table}
	\end{center}
\end{table}

\newpage

\subsection{Kepler from Kepler Exploration Robotics}

\begin{table}[ht!]
	\begin{center}
		\begin{tabular}{  p{15cm} }
			\toprule
			Kepler from \href{https://www.gotokepler.com/home}{Kepler Exploration Robotics} \hfill  \includegraphics[scale=0.1]{china.eps} \\ 
			
			\cmidrule(lr){1-1}
			
			\begin{wrapfigure}{l}{0.25\textwidth}
				\centering
				\includegraphics[scale=0.3]{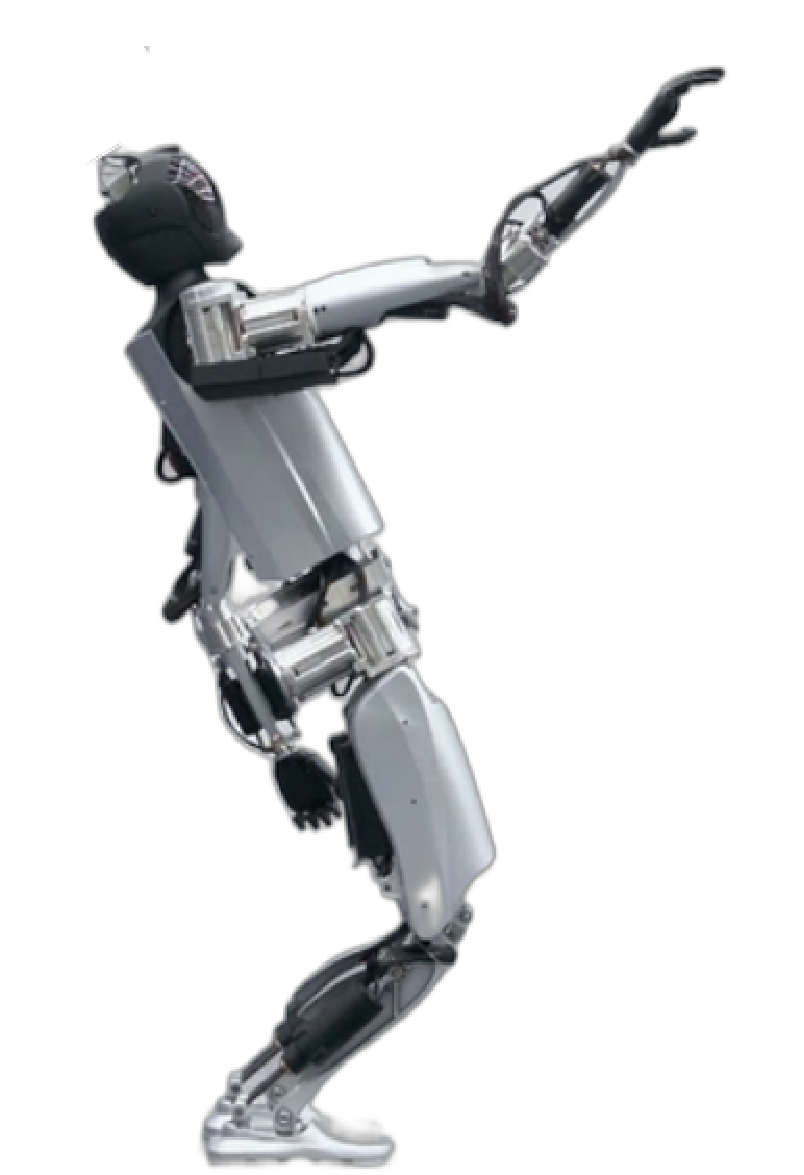}
			\end{wrapfigure}
			
			\textbf{Robotics background} (0.2) Despite the fact that Kepler was showcased at CES 2024~\cite{PRNewswire_kepler}, there is few information available regarding the company itself. However, in this \href{https://www.youtube.com/watch?v=uLNvRVoLlHA}{video} from {TheAIGRID} one can see Kepler walking and manipulating some objects.
			
			\textbf{Teleoperation} (0) Kepler Exploration Robotics did not mention teleoperation as a tool to either train the humanoid or a possibility to take over in case the humanoid is stucked.
			
			\textbf{Modularity} (0) Kepler is selling as a full humanoid robot.
			
			\textbf{Dexterous hand} (1) Kepler Exploration Robotics designed a hand with 12 DoF. This \href{https://www.youtube.com/watch?v=uLNvRVoLlHA}{video} shows Kepler manipulating some objects.
			
			\textbf{Task planning} (0.2) Kepler Exploration Robotics has equipped the humanoid robot with a cloud-based multimodal large model while simultaneously deploying a smaller, industry-specific model on the robot itself for quicker response times~\cite{PRNewswire_kepler}. It means that engineers are probably working on task planning but not demonstrated it.
			
			\textbf{AI} (0.4) Kepler Exploration Robotics is developing its own AI stack named {NEBULA} enabling the robot to interact with the surrounding environment in real time. The humanoid is equipped with a processor providing 100 TOPS of computing performance into the NEBULA system composed of visual recognition, visual SLAM (Simultaneous Localization and Mapping), multimodal interaction, and hand-eye coordination~\cite{PRNewswire_kepler}. 
			
			\textbf{Walking gait robustness} (0.6) This \href{https://www.youtube.com/watch?v=hnqnom4KUcE}{video}  showed Kepler and its balancing ability but the company did not disclose videos demonstrating dynamic walking capabilities of Kepler.
			
			\textbf{Market/Pilot} (0.2) According to company website,Warehousing, smart inspection, automated product lines, outdoor tasks and high risk operations are the targeted markets. Different versions K1,S1 and D1) of Kepler will be built to address these different markets. Kepler Exploration Robotics offers an open development platform, inviting developers and integrators to create innovative solutions that leverage the robot’s advanced capabilities.
			
			\\ \bottomrule
			
			\begin{multicols}{2}
				\begin{itemize}
					\item Height: 175cm
					\item payload: unknown
					\item weight 85kg
					\item runtime: unknown
					\item walking speed 1-2m/s
					\item Degrees of Freedom: 40 (Hand 12 DoF) 
					\item Perception: sensor set Infrared binocular 3D Camera
					\item Target Price: Around US\$ 30K
				\end{itemize}
			\end{multicols}
			\\ \bottomrule
			
		\end{tabular}
		\caption{Short introduction of Kepler from Kepler Exploration Robotics}
		\label{kepler_table}
	\end{center}
\end{table}

\newpage

\subsection{CL-1 from LimX Dynamics}

\begin{table}[ht!]
	\begin{center}
		\begin{tabular}{  p{15cm} }
			\toprule
			{CL-1} Biped from \href{http://www.limxdynamics.com/en}{{LIMX} Dynamics} \hfill  \includegraphics[scale=0.1]{china.eps} \\ 
			
			\cmidrule(lr){1-1}
			
			\begin{wrapfigure}{l}{0.25\textwidth}
				\centering
				\includegraphics[scale=0.2]{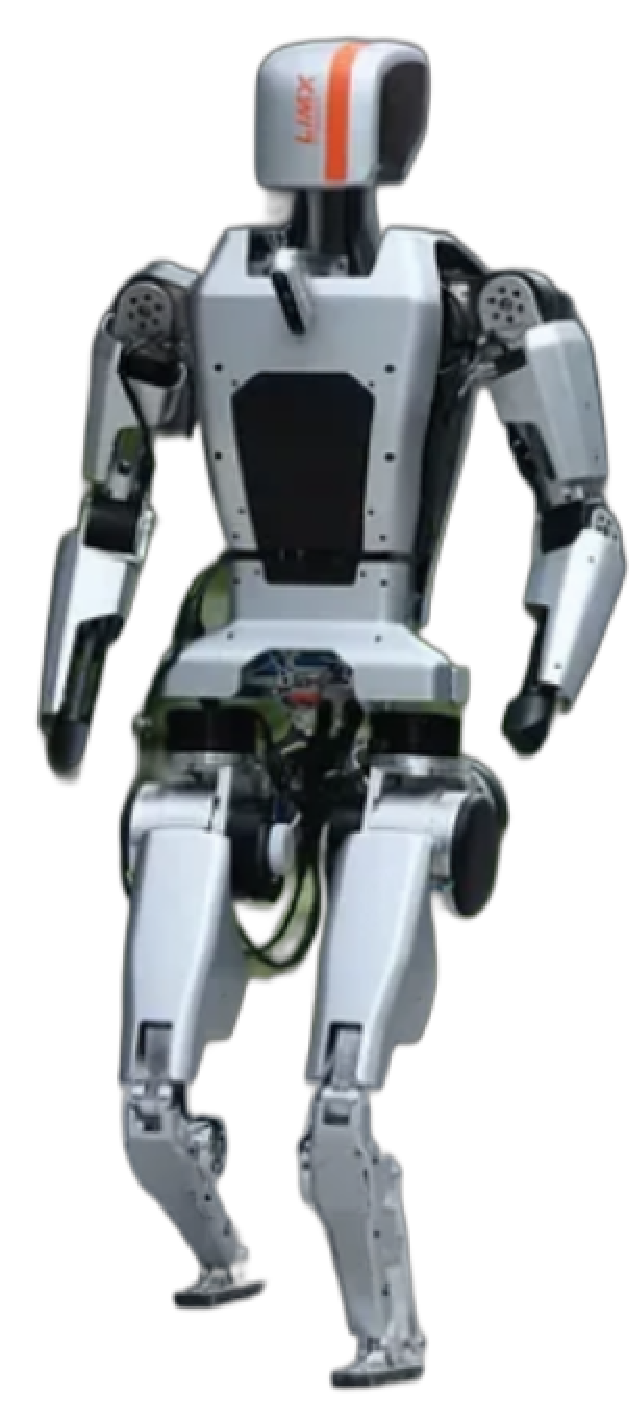}
			\end{wrapfigure}
			
			\textbf{Robotics background} (0.5):  LimX Dynamics has been founded in 2022 and is based in Shenzhen. They already propose X1 a quadruped robot and W1 a wheeled quadruped robot. {CL-1} the humanoid robot was showcased in December 2023~\cite{limXDynamicsCL-1}. Like Unitree - see Table~\ref{Unitree_table} - the development of {CL-1} relies on the algorithms and actuators developed for the quadurped robots.  Founded by a group of robotics scientists, the startup has already raised 200 million yuan (US\$27.5 million) in angel and pre-A financing. Along with the hiring of Zhang Li, The former \href{https://www.weride.ai/}{WeRide} chief operating officer, the startup is also appointing Dr. Jia Pan, a tenured associate professor at the University of Hong Kong, as its chief scientist~\cite{techcrunch_limx}.
			
			\textbf{Teleoperation} (0.5): according to this \href{https://www.youtube.com/watch?v=2dmjzMv-y-M}{video}, {LIMX} Dynamics uses tleoperation to train their humanoid.
			
			\textbf{Modularity} (0): {CL-1} Biped is selling a walking humanoid robot only.
			
			\textbf{Dexterous hand} (0): {CL-1} is not equipped with hands yet.
			
			\textbf{Task planning} (0): {LIMX} Dynamics did not mention whether they are investigating task planning or not.
			
			\textbf{AI} (0.4):{LIMX} Dynamics relies on AI for perception and objects classification and the humanoid is not equipped with hands so far. 
			
			\textbf{Walking gait robustness} (0.6): This \href{https://www.youtube.com/watch?v=sihIDeJ4Hmk}{video}  showed the {CL-1} humanoid robot is able to walk on different types of surface and even climb stairs. This \href{https://www.youtube.com/watch?v=UpNid_rWDnI}{video} demonstrates the expertise of the team in its capability to develop a very robust and dynamic gait for a biped, without upper body, using reinforcement learning.
				
			\textbf{Market/Pilot} (0): LimX Dynamics humanoid robots will be progressively deployed in both B2B and B2C applications, focusing on hazardous scenarios, high-end services, automobile manufacturing, and in-home services~\cite{limXDynamicsCL-1}. At this moment, the humanoid is still in development.

			\\ \bottomrule
			
			\begin{multicols}{2}
				\begin{itemize}
					\item Height: Not disclosed
					\item Payload: Not disclosed
					\item Weight: Not disclosed
					\item Runtime: Not disclosed
					\item Walking speed: 1-2m/s
					\item Degrees of Freedom:18 (no hand)
					\item Perception: Depth Cameras
					\item Target Price: Not disclosed
				\end{itemize}
			\end{multicols}
			\\ \bottomrule
			
		\end{tabular}
		\caption{Short introduction of {CL-1} from LimX Dynamics}
		\label{limx_table}
	\end{center}
\end{table}

\end{document}